\documentstyle [12pt,epsf] {article}
\def\dim{dimension}
\def\ss{Scherk--Schwarz\ }

\newcommand{\sect}[1]{ \section{#1} \setcounter{equation}{0} }

\newcommand{\subsect}{\subsection}
\newcommand{\req}[1]{(\ref{#1})}
\newcommand{\nwc}{\newcommand}
\nwc{\hyp} {\hyphenation}

\def\bfone{\relax{\rm 1\kern-.35em 1}}
\def\inbar{\vrule height1.5ex width.4pt depth0pt}
\def\IC{\relax\,\hbox{$\inbar\kern-.3em{\mss C}$}}
\def\ID{\relax{\rm I\kern-.18em D}}
\def\IF{\relax{\rm I\kern-.18em F}}
\def\IH{\relax{\rm I\kern-.18em H}}
\def\II{\relax{\rm I\kern-.17em I}}
\def\IN{\relax{\rm I\kern-.18em N}}
\def\IP{\relax{\rm I\kern-.18em P}}
\def\IQ{\relax\,\hbox{$\inbar\kern-.3em{\rm Q}$}}
\def\IZ{{\bf Z}}
\def\Li{{\cal L}i}
\nwc{\ov}  {\overline}
\nwc{\be}  {\begin{equation}}
\nwc{\ee}  {\end{equation}}
\nwc{\ba}  {\begin{array}}
\nwc{\ea}  {\end{array}}
\nwc{\bdm} {\begin{displaymath}}
\nwc{\edm} {\end{displaymath}}
\nwc{\bea} {\be\ba{lcl}}
\nwc{\eea} {\ea\ee}
\nwc{\bda} {\bdm\ba{lcl}}
\nwc{\eda} {\ea\edm}
\nwc{\bc}  {\begin{center}}
\nwc{\ec}  {\end{center}}
\nwc{\ds}  {\displaystyle}

\nwc{\nn} {\nonumber}
\nwc{\nnn} {\nonumber \vspace{.2cm} \\ }
\nwc{\ra}{\rightarrow}
\nwc{\lra}{\longrightarrow}
\nwc{\p} {\partial}

\nwc{\h} {\frac{1}{2}}
\nwc{\fc} {\frac}
\nwc{\lf}{\left}
\nwc{\ri}{\right}
\def\Rc{{\cal R}}
\def\Fc{{\cal F}}
\def\Cc{{\cal C}}

\def\Hc{{\cal H}}
\def\Oc{{\cal O}}
\def\Ec{{\cal E}}
\def\Vc{{\cal V}}

\if@twoside\oddsidemargin23mm
\else\oddsidemargin23mm\evensidemargin23mm\marginparwidth23mm\fi%
\begin{document}

\begin{titlepage}
\begin{flushright}  
{hep--th/0108183}\\
{BONN-TH-01-21}\\   
{NUB--3218}\\ 
{August 2001}\\
\end{flushright}  

\vskip1cm
\begin{center} 
{\large\sc Divergences  in Kaluza--Klein Models and their}\\
\vspace{0.4cm}
{\large\sc String Regularization}\\
\vspace{1.5cm}
{\large D.M. Ghilencea$^a$, H.P. Nilles$^a$}
\ and\ \ {\large S. Stieberger$^b$}
\bigskip \\[0pt] 
\vspace{1cm} 
{\sl $^a$ Physikalisches Institut der Universit\"at Bonn,} \\
{\sl Nussallee 12, 53115 Bonn, Germany}\\[2mm]
{\tt Email: Dumitru@th.physik.uni-bonn.de, Nilles@th.physik.uni-bonn.de}\\
\vspace{1.0cm}
{\sl $^b$ Department of Physics, Northeastern University, 
Boston, MA 02115, U.S.A.}\\[2mm]
{\tt Email: stieberg@infeld.physics.neu.edu}\\
\bigskip
\end{center}

\vspace{3cm}
\begin{center}
{\bf Abstract}\\
\vspace{0.5cm}
\end{center}
Effective field theories with (large) extra dimensions
are studied within a physical regularization scheme 
provided by string theory.
Explicit string calculations then allow us to consistently 
analyze the ultraviolet sensitivity of Kaluza--Klein theories 
in the presence or absence of low energy supersymmetry.

\vfill
\thispagestyle{empty}
\end{titlepage}


\newpage
{\it\tableofcontents}

\newpage
\sect{Introduction}

The main motivation to consider generalizations of the Standard Model
(SM) of strong and electroweak interactions is the hierarchy problem:
the instability of the weak scale 
$M_w \approx G_F^{-1/2}\approx 300$ GeV in the presence
of higher scales, like $M_{Planck}\approx 10^{19}$ GeV connected to
the existence of gravitational interactions.
The hierarchy problem is linked to a power law sensitivity of the low
energy effective field theory to the ultraviolet energy region. In the 
standard model such a sensitivity (for field dependent
quantities\footnote{In addition there is a quartic
divergence of the vacuum energy 
(cosmological constant).}) manifests itself in
the quadratic divergence of the (mass)$^2$ of the scalar Higgs
particle.

Supersymmetry has been suggested as a possible solution to this
problem. In the supersymmetric extension  of the standard model with 
softly broken supersymmetry, no power-like divergences appear (with
the exception of the cosmological constant) to all orders in
perturbation theory. The Higgs (mass)$^2$  still shows a logarithmic
sensitivity to the ultraviolet energy scale, but this is consistent
with a stabilization of the weak scale.

More recently, with the thorough discussion of effective field
theories with (large) extra dimensions the hierarchy problem 
reappears again and it has been claimed in the literature that in 
specific cases the weak scale does not show a power-like 
ultraviolet behaviour even in the absence of low-energy
supersymmetry. These claims are very surprising  since the considered
effective field theories are non-renormalizable and therefore contain
many more sources of power-like divergences. While in the standard
model (as a renormalizable $D=4$ field theory) the (mass)$^2$ of a scalar
particle is the only source of a quadratic divergence (for field
dependent quantities) in higher dimensional theories we might find
power-law sensitivity for gauge couplings, Yukawa couplings as well as 
new higher dimensional operators. It is very difficult to analyze
these questions within a naive low energy effective field theory
approach (well suited for  renormalizable theories). This is
especially true if the discussion relies exclusively on the
consideration of one-loop calculations since the (infinitely many) new
counterterms might manifest themselves only in higher loop
contributions to the effective Higgs mass as a power-like divergence.
A careful examination of this question is thus needed in the
framework of these higher dimensional theories, especially since the
finiteness of one loop results might be obtained by a specific
choice of the regularization procedure.
 
In the present paper we provide a physical regularization scheme for
such theories via the embedding in string theory. The motivation is
two-fold.
First of all we believe that a meaningful description of these extra
dimensional (non-renormalizable) theories requires a more fundamental
theory at the high scale (for which string theory is a candidate).
Secondly, in a consistent (finite) string theory perturbative 
calculations give a finite result and we do not have to deal 
with the often arguable choice of a field theory regularization
scheme. String theory provides us with a field theory ultraviolet
cut-off, the string tension $M_{string}=\alpha'^{-1/2}$ with 
$\alpha'$ the slope parameter.
The ultraviolet sensitivity of a physical parameter (like the Higgs
mass) can be read off from the dependence on $M_{string}$, taking
into account the appropriate sensitivity of all quantities appearing
in that expression (like e.g. a power-like sensitivity of the coupling
constants). Such a discussion is conceptually simpler and more
powerful than a naive field theory consideration as it is independent
of specifically chosen regularization schemes. Of course, the low
energy field theory description can be obtained in the limit
$M_{string}\ra \infty$ ($\alpha'\ra 0$, zero slope limit) where the
power-like sensitivity with respect to $M_{string}$ manifests itself 
in an ultraviolet, power-like divergence.

The first part of the paper addresses the scalar potential in
effective  field theories with an additional compact dimension and
the need for a string regularization scheme. This is  provided in 
section \ref{stringreg}, where the link with string theory 
is made explicit. The second part of the paper
provides details of the calculation of the vacuum 
energy, one--loop gauge and Yukawa couplings in a  significant 
class of string models.

\sect{Field theory calculation of the scalar potential}\label{scalarpot}
\subsect{The scalar potential}\label{sp}
In this section we investigate\footnote{For early research  see e.g. 
refs. \cite{oldcosmo}.}
the scalar potential  in a class of models  with Kaluza Klein towers of states 
associated with one additional spatial dimension.  This class of
models \cite{antoniadis,Delgado,Hall,Barbieri,Quiros}
received renewed interest recently due to their 
very interesting phenomenological consequences. The introductory part
of this analysis was already published in \cite{dmg} and we only review it
for continuity of our presentation.

The aforementioned models usually consider as a starting point
a higher dimensional theory (e.g. 5D) with Scherk Schwarz \cite{ss} 
and/or orbifold mechanism for supersymmetry breaking. 
The presence of an additional (compact) dimension induces in the 4D
(boundary) effective  theory a tower of Kaluza Klein modes which, depending on
the particular model can be associated with various states of the
low energy spectrum. These modes fall in N=2 multiplets, consequence
of the enhanced supersymmetry in the 5D ``bulk'' broken to N=0 on the 4D
boundary by an orbifold-like compactification 
(for example $S_1/(Z_2\times Z_2')$  \cite{Barbieri}).
Yukawa interactions are localized on the 4D ``boundary'' and induce
(together with the gauge part) 
corrections to the scalar potential of these models. 
These will be addressed in the following. 
Gauge corrections will be considered elsewhere \cite{toappear}. 
For a detailed description of generic models see for example 
 \cite{Delgado,Barbieri}.

To begin with,  it is instructive for our purposes to explicitly 
present the contribution of one Kaluza Klein bosonic or fermionic mode of mass
$m_k$ to the scalar potential, computed in a one-loop expansion of the
latter and in all orders in perturbation theory.
We denote this contribution $\Vc_k^o$ and we use a cut-off regularization
method, with $\Lambda$  the momentum cut-off of the {\it effective}
field theory (EFT). For review and applications
of this regularization method to the scalar potential see \cite{Sher}. 
Every regularization scheme has its own shortcomings, and our
choice is  motivated on physical grounds. Unlike other regularization 
methods, it is appropriate for displaying some phenomenological 
implications of what is called \cite{Hall} 
``Kaluza Klein regularization'' 
which cannot be ``seen'' in  dimensional regularization for example. 
Other, more elegant regularization schemes are possible,  
but the present choice is more suitable to understand
the  need for and link with a full string  calculation/regularization 
(see later). Thus $\Lambda$ is the scale where new physics (string physics)
comes in. We  have
\begin{eqnarray}\label{Vn}
\Vc_k^o&=&\frac{1}{(2\pi)^4}\int d^4 p\, \ln(1+m_k^2/p^2)
\nonumber\\
&=&\frac{2\pi^2}{(2\pi)^4}
\left\{\frac{1}{4}\left[
m_k^2 \Lambda^2 -m_k^4\ln\left(1+\frac{\Lambda^2}{m_k^2}\right)
+\Lambda^4\ln\left(1+\frac{m_k^2}{\Lambda^2}\right)
\right]\right\}
\end{eqnarray}
As it is well-known, the contribution to the scalar potential
of one (Kaluza Klein) state contains  quadratic and logarithmic
terms, the first and second term in (\ref{Vn}). 

We consider now the overall contribution to the one loop 
expansion of the scalar  potential  of towers of Kaluza Klein (KK)
modes which, following ref.~\cite{Barbieri} are
associated with the top quark due to its larger Yukawa coupling.
The scalar  potential for the Higgs field 
$\phi$   has  the following generic structure after  taking into 
account individual  bosonic and fermionic Kaluza Klein 
contributions  (\ref{Vn}), 
\begin{equation}\label{potential1}
\Vc(\phi)=\frac{1}{2} Tr \sum_{k=-l}^{l}\int \frac{d^4 p}{(2\pi)^4}
\ln\frac{p^2+m_{B_k}^2(\phi)}{p^2+m_{F_k}^2(\phi)}
\end{equation}
with the Trace taken over the top hypermultiplet of fixed k level
contributing a factor of $(4 N_c)$ where $N_c$ is the number of
colors. 

In this analysis we will  sum over the whole Kaluza-Klein tower
of states, $l\ra \infty$.
For this purpose we first consider the 
case of an arbitrarily fixed number ($l$) of Kaluza-Klein modes
that we sum over and arbitrarily fixed momentum cut-off $\Lambda$. 
Thus there are two  cut-off's corresponding to the compact  and
non-compact sectors.  We  then take the 
limit $l\gg (\Lambda \Rc)/2$. Further we  consider the limit 
$l\ra \infty$ and then $\Lambda\ra \infty$ to recover the
case  of ``Kaluza Klein regularization''. 
For a more suitable regularization for models with compact and non-compact
sectors  see \cite{steph}. 

The field dependent boson and fermion masses in (\ref{potential1})
are usually given by 
\begin{equation}\label{fermion}
m_{F_k}(\phi)=\frac{2 k}{{\cal R}}+
m_t(\phi)=\frac{2}{{\cal R}}(k+\omega),\,\,\,\,\,\,
\omega=\frac{m_t(\phi){\cal R}}{2}
\end{equation}
and
\begin{equation}\label{boson}
m_{B_k}(\phi)=\frac{2 k+1}{{\cal R}}+
m_t(\phi)=\frac{2}{{\cal R}}(k+\omega'),\,\,\,\,\,
\omega'=\omega+\frac{1}{2}
\end{equation}
where we follow \cite{Barbieri} and  consider any integer 
values of $k$ between $-\infty<k<\infty$. These mass formulas are 
generic for the states of the Kaluza Klein tower and depend 
(via $\omega$ and $\omega'$) on the boundary conditions 
one chooses for fermions/bosons. 
Other cases for the  mass assignment exist in the
literature  \cite{antoniadis,Delgado} which we review 
in section \ref{case2}. There, the mass (squared) of KK states is 
in addition shifted by an equal amount for both bosons and fermions. 

The mass splitting between fermions and bosons, 
following the  breaking of supersymmetry 
via an orbifold or Scherk-Schwarz mechanism,  is then encoded in the
difference $\omega'-\omega$ which for this case is positive, 
eq.(\ref{boson}). 
The sign of the scalar potential
and also of  its second derivative  (at $\phi=0$) will then depend
on the sign of this difference (see eq.(\ref{potential1})), 
with implications for the existence of the (radiatively induced)
electroweak symmetry breaking mechanism.
The existence of this  mechanism is then 
traced back to the type of boundary 
conditions one chooses for 
the fermions and for the bosons respectively. 

From equations (\ref{potential1}), (\ref{fermion}) and  (\ref{boson})
we find 
\begin{equation}\label{potential2}
\Vc(\phi)=\frac{\eta}{{\cal R}^4}
\sum_{k=-l}^{l}\int_{0}^{\overline\Lambda} {d \rho}\, 2\, \rho^3 
\ln\frac{\rho^2+\pi^2( k+\omega')^2}
{\rho^2+\pi^2( k+\omega)^2},\,\,\,\,\,\,\,\,\,\,
{\overline\Lambda}=\frac{\pi {\cal R} \Lambda}{2},\,\,\,\,\,\,\,\,\,\,
\eta=\frac{4 N_c}{2 \pi^6}
\end{equation}
where $\Lambda$ is the momentum cut-off of the loop-integral 
in (\ref{potential1}) and $l$ the number of states in the tower.
Intuitively, in an effective field theory (of fixed cut-off $\Lambda$)
the number of Kaluza-Klein states should be restricted 
(unless a symmetry prevents us from doing so\footnote{
We return to this point in Section \ref{stringreg}.}) 
to those states
whose mass is smaller than the momentum cut-off of the loop integral,
condition which would lead to 
\begin{equation}
m_{B_k, F_k}\approx \frac{2 k}{{\cal R}} \leq \Lambda \,\,\,\,\,\,
\Rightarrow \,\,\,\,\,\, {\overline\Lambda}\approx \pi l 
\end{equation}
Here $l$ stands for  the Kaluza Klein state of the largest 
mass. Performing the sum over the tower of 
Kaluza-Klein states  under the momentum integral  
eqs.(\ref{potential1}), (\ref{potential2}) gives (see Appendix A)
\begin{equation}\label{sumtol}
\sum_{k=-l}^{l}\ln\frac{\rho^2+\pi^2( k+\omega')^2}
{\rho^2+\pi^2( k+\omega)^2}={\cal Z}_0+{\cal Z}_1+{\cal Z}_2
\end{equation}
with the notation
\begin{equation}\label{z0}
{\cal Z}_0=\ln\frac{\cosh(2 \rho)-\cos(2\pi\omega')}
{\cosh(2 \rho)-\cos(2\pi\omega)}
\end{equation}
and
\begin{equation}\label{z1}
{\cal Z}_1=
\ln\frac{[\rho^2+\pi^2( l\pm\omega')^2]_*}
{[\rho^2+\pi^2( l\pm\omega)^2]_*}
\end{equation}
and finally
\begin{equation}\label{z2}
{\cal Z}_2=
\ln\frac{[\Gamma(l \pm \omega'\pm i\rho/\pi)]_*}
{[\Gamma(l \pm \omega\pm i\rho/\pi)]_*}
\end{equation}
The symbol $[\rho^2+\pi^2( l\pm\omega)^2]_*$ stands for a product 
of the quantity within the  brackets with all 
possible combinations of plus and minus signs.
Accordingly, the potential can  be written
($\phi$ dependence hidden in $\omega$ and $\omega'$)
\begin{equation}
\Vc(\phi)=\Vc_0(\phi)+\Vc_1(\phi)+\Vc_2(\phi)
\end{equation}
where we have
\begin{equation}\label{pot0}
\Vc_0(\phi)=\frac{\eta}{{\cal R}^4}
\int_{0}^{\overline\Lambda} {d \rho}\, 2\, \rho^3 {\cal Z}_0=
 \frac{\eta}{{\cal R}^4}
\int_{0}^{\overline\Lambda} {d \rho}\, 2\, \rho^3 
\ln\frac{\cosh(2 \rho)-\cos(2\pi\omega')}
{\cosh(2 \rho)-\cos(2\pi\omega)}
\end{equation}
and
\begin{equation}\label{pot1}
\Vc_1(\phi)=\frac{\eta}{{\cal R}^4}
\int_{0}^{\overline\Lambda} {d \rho}\, 2\, \rho^3 {\cal Z}_1
=
\frac{\eta}{{\cal R}^4}
\int_{0}^{\overline\Lambda} {d \rho}\, 2\, \rho^3 
\ln\frac{\rho^2+\pi^2( l+\omega')^2}
{\rho^2+\pi^2( l+\omega)^2}+(l \rightarrow -l)
\end{equation}
and finally
\begin{equation}\label{pot2}
\Vc_2(\phi)= \frac{\eta}{{\cal R}^4}
\int_{0}^{\overline\Lambda} {d \rho}\, 2\, \rho^3 {\cal Z}_2
=\frac{\eta}{{\cal R}^4}
\int_{0}^{\overline\Lambda} {d \rho}\, 2\, \rho^3 
\ln\frac{[\Gamma(l \pm \omega'\pm i\rho/\pi)]_*}
{[\Gamma(l \pm \omega\pm i\rho/\pi)]_*}
\end{equation}
For later reference it is useful to remind that the 
numerators ($\omega'$ dependent) of the integrands of 
$\Vc_0$, $\Vc_1$, $\Vc_2$ correspond to bosonic degrees of freedom, 
while the denominators  correspond to the contribution ($\omega$
dependent) of the fermions. 

In the following we analyse each of the three contributions to
the scalar potential $\Vc$. 
$\Vc_0$ is part of the scalar potential corresponding to the result of 
reference \cite{Barbieri} which is finite and ultraviolet insensitive,
as we discuss below. The expression of $\Vc_0$ 
 simply corresponds to summing  over an infinite number of Kaluza
Klein states in eq.(\ref{potential1})  
and is therefore Kaluza-Klein-level independent. It is considered 
\cite{antoniadis,Delgado,Hall,Barbieri} that this is the 
Kaluza-Klein regularized part of the scalar potential. 

The contributions $\Vc_1$ and $\Vc_2$ to the scalar potential 
are each   vanishing in the limit of an
infinite number of Kaluza-Klein states. Indeed
\begin{equation}
\Vc_1(l\rightarrow \infty)=0
\end{equation}
and 
\begin{equation}
\Vc_2(l\rightarrow \infty)=0
\end{equation}
while keeping $\overline\Lambda$ fixed, consequence of the vanishing
of the integrands of $\Vc_1$ and $\Vc_2$ respectively. 
In this limit only $\Vc_0$ will contribute to the scalar potential,
\begin{equation}
\Vc(l\ra \infty)=\Vc_0
\end{equation}
to give the result of \cite{Barbieri}.
We  emphasize that it is instructive for our purposes to 
explicitly compute $\Vc_1$ and $\Vc_2$  in the general case of finite
$l$ and then take the limit $l\rightarrow \infty$ ($\overline\Lambda$ fixed)
to understand its physical implications.
This procedure/limit will clarify the role
of supersymmetry  in the cancellation of individual  quadratic and
logarithmic terms of the  potential.  Note that this limit in
eq.(\ref{Vn}) means  that its first two terms (quadratic and
logarithmic)  are not manifest in the final result for $\Vc$ 
obtained\footnote{One can see that the 
limit of very large $m_k$ (large KK level) 
$m_k\gg\Lambda$ in eq.(\ref{Vn})
means that the first and second terms cancel in the 
fermionic or bosonic part
only to leave $\Lambda^4$ terms (to be canceled due to equal
bosonic/fermionic degrees of freedom.).
We return to this observation later in the text.}
after summing up infinitely many contributions to the scalar 
potential. For clarity it is perhaps worth anticipating some of our
conclusions,  that in the limits of either $l\gg (\Lambda \Rc)/2$
or  of summing the whole KK tower
the quadratic and logarithmic divergences in $\Vc$ are absent
in the bosonic sector alone and the same mechanism applies for the fermionic
sector, {\it without} the need of supersymmetry.

\subsubsection{The limits $l\gg (\Lambda \Rc)/2$ and $l\ra \infty$
(``Kaluza Klein regularization'')}\label{KKlimit}
After performing the 
integral  for $\Vc_0$,  one finds the following result\footnote{
$\gamma(\alpha,x)$ is the incomplete Gamma function \cite{gradshteyn}.}.
\begin{equation}
\Vc_0(\phi)=-\frac{N_c}{2 \pi^6 {\cal R}^4} 
\left\{\frac{1}{2}
\sum_{n=1}^{\infty} 
\frac{e^{2 i \pi n \omega'}}{n^5}\,
\gamma(4, 2 n {\overline\Lambda} )+ (\omega'\ra -\omega')
\right\}-(\omega'\ra \omega)\label{omegam}
\end{equation}
For $\omega'=\omega+1/2$ (eq.\ref{boson}) the above relation 
reduces to
\begin{equation}\label{el}
\Vc_0(\phi)= \frac{3\, \eta}{{\cal R}^4}\sum_{k=0}^{\infty}
\frac{\cos(2 \pi\omega(2k+1))}
{(2k+1)^5}\left\{1-e^{-2{\overline\Lambda}(2k+1)}
\sum_{m=0}^{3}\frac{(2 {\overline\Lambda}(2k+1))^m}{m!}\right\}
\end{equation}
where the factor within curly braces is equal to unity in the limit of
large\footnote{In
the case we set $\omega=0$ we find  that the potential is equal to
\begin{equation}
\Vc_0(\omega=0)=\frac{93 (4N_c)}{64 \pi^6 {\cal R}^4}\zeta(5)
\end{equation}
and is (up to trace factor of $(4N_c)$)
the result of \cite{Dimopoulos} in our notations/assignment 
conventions of the KK tower. } ${\overline\Lambda}$. 
The sign of the potential 
(fixed by the choice of boundary conditions for the bosons/fermions) 
is appropriate for triggering the 
electroweak symmetry breaking.  Further analysis of this potential 
and phenomenological implications have been discussed \cite{Barbieri}.

The result of integrating $\Vc_1(\phi)$ of eq.(\ref{pot1}) 
is given, irrespective of the 
dependence $m_t(\phi)$, by the following expression
\begin{equation}\label{sum}
\Vc_1={\overline \Vc}_1(\omega')-{{\overline \Vc}}_1(\omega)
\end{equation}
With the exception of some $\Lambda^4$
terms  canceled in eq.(\ref{sum}) between bosonic and fermionic contributions 
(due to equal number of bosonic/fermionic degrees of freedom), 
${{\overline \Vc}}_1(\omega')$ represents 
the sole contribution of the 
bosonic ($\omega'$ dependent) sector, while ${{\overline \Vc}}_1(\omega)$
is that of the  fermionic sector. We have
\begin{eqnarray}\label{bosonic}
{{\overline \Vc}}_1(\omega')&=&\frac{\eta}{2 {\cal R}^4}
\left\{\pi^2(l+\omega')^2 {\overline\Lambda}^2
-\pi^4(l+\omega')^4\ln\left[1+\frac{{\overline\Lambda}^2}
{\pi^2(l+\omega')^2}\right]
+{\overline\Lambda}^4\ln\left(\pi^2(l+\omega')^2
+{\overline\Lambda}^2\right)\right\}\nonumber\\
&+&(l\rightarrow -l)
\end{eqnarray}
As expected, both ${{\overline \Vc}}_1(\omega')$ and 
${{\overline \Vc}}_1(\omega)$ have quadratic and logarithmic divergences
and these are not canceled between bosons and fermions for any finite 
summation over the KK tower of states.
The absence of such ultraviolet  terms in the final 
result for $\Vc$ in the limit of summing over an infinite tower of
states is sometimes considered as 
due to the ``soft nature'' of  breaking supersymmetry
by a Scherk-Schwarz mechanism. This  would apparently ensure a 
cancellation between 
bosonic and fermionic contributions not only at the level
of quartic divergences ($\Lambda^4$), but also  at the level of quadratic
($\Lambda^2$) and logarithmic ($\log\Lambda$) terms.
This is not necessarily true, for reasons that we discuss below.

We consider  in the following the limit of  large $l$ and  
fixed  $\Lambda$,  
$\pi l\gg\overline\Lambda\equiv\pi\Lambda {\cal R}/2$. 
This seems a strange limit to take in a 4D effective field theory, 
of considering KK modes of mass $l/\Rc\gg \Lambda$.
In this case  the logarithm in the  second term of the bosonic term 
${{\overline \Vc}}_1(\omega')$ 
can be expanded in a (rapidly convergent) 
power series, with the first term in the expansion
to cancel the quadratic divergence ${\overline\Lambda}^2$ 
of the first term in ${{\overline \Vc}}_1$; more explicitly, the second term 
in (\ref{bosonic}) denoted as $A(l,\omega')$ is given by
\begin{equation}\label{xpansion}
A(l,\omega')\approx 
\frac{1}{2}\frac{\eta}{{\cal R}^4}\left\{
-\pi^2(l+\omega')^2 {\overline\Lambda}^2
+\frac{1}{2}{\overline\Lambda}^4-\frac{1}{3}
\frac{{\overline\Lambda}^6}{\pi^2(l+\omega')^2}+\cdots\right\}
\end{equation}
and its quadratic term\footnote{The  $\Lambda^4$ term in the expansion 
(\ref{xpansion}) 
is canceled in $\Vc_1$ after including both bosonic and fermionic
contributions.} cancels the first term in ${{\overline \Vc}}_1(\omega')$.
This  cancellation takes place separately  for the bosons and for
the fermions. The last term (and higher) in (\ref{xpansion}) are
suppressed when $l\gg (\Lambda \Rc)/2$.
The interpretation of this observation is that states 
of mass  larger  than the momentum cut-off scale 
$\Lambda$ of the loop  integral ($2 l/{\cal R} \gg \Lambda$) are required to
cancel  individual quadratic divergences of Kaluza-Klein states.
This situation has an analogue in eq.(\ref{Vn}) where we would take the limit 
$m_k\gg \Lambda$ with $m_k$  the mass of a Kaluza Klein
bosonic or fermionic state. There the quadratic 
term in (\ref{Vn}) would be canceled by the first term in the (convergent) 
expansion of the logarithmic contribution.

Summing over an infinite tower of Kaluza Klein states
corresponds to the limit $l\rightarrow \infty$ 
in which case one interchanges the infinite Kaluza Klein sum
and the momentum integral. While this is  possible\footnote{We 
discuss this later in the text.}, 
it  includes modes of mass larger
than the momentum cut-off of the loop integral also 
intended to be the cut-off of our effective field theory. 
When $\l\ra \infty$,
the quadratic and logarithmic dependence in the bosonic part
${{\overline \Vc}}_1(\omega')$ of (\ref{bosonic})
disappears  (higher order terms in (\ref{xpansion}) vanish) and a 
similar mechanism separately applies for the fermionic part. Indeed
\begin{equation}\label{v1limit}
\lim_{l\rightarrow \infty}
\frac{1}{2}\frac{\eta}{{\cal R}^4}
\left\{\pi^2(l+\omega')^2{\overline\Lambda}^2
-\pi^4(l+\omega')^4\ln\left[1+\frac{{\overline\Lambda}^2}
{\pi^2(l+\omega')^2}\right]
+(l\rightarrow -l)\right\}=\frac{1}{2}\frac{\eta}{{\cal R}^4}\,
{\overline\Lambda}^4
\end{equation}
where the r.h.s. 
${\overline\Lambda}^4$ is canceled between bosons and fermions
due to their equal number ensured by the initial (now broken)
supersymmetry.

The expression of $\Vc_2$, eq.(\ref{pot2}) can be integrated in the
approximation  of large modulus of $l+i{\overline\Lambda}$ which 
does not restrict the relative values of $l$, ${\overline\Lambda}$ 
and will thus not affect the result we obtain for $\Vc_2$ in the
limit $l\gg {\overline\Lambda}$. The approximation we use 
for the integrand of $\Vc_2$ is detailed in Appendix~B.
After some algebra the result we obtain for $\Vc_2$, eq.(\ref{pot2})
may be written as
\begin{eqnarray}\label{v2}
\Vc_2& \approx & \frac{\eta}{{\cal R}^4}
\left\{   \pi^2 g(l,\omega')(l+\omega')
\left[ {\overline\Lambda}^2-\pi^2(l+\omega')^2
\ln\left(1+\frac{ {\overline\Lambda}^2 }{\pi^2(l+\omega')^2}\right)
\right]
+(\omega'\rightarrow -\omega')\right\}\nonumber\\
&+&
\frac{\eta}{{\cal R}^4}\frac{ \omega'{\overline\Lambda}^4}{2}
\left\{\ln\frac{{\overline\Lambda}^2+\pi^2( l+\omega')^2}
{{\overline\Lambda}^2+\pi^2( l-\omega')^2}\right\}
-\frac{\eta}{{\cal R}^4}
\frac{4{\overline\Lambda}^5}{5 \pi}
\left\{\arctan\frac{\overline\Lambda}{\pi(l-\omega')}
+\arctan\frac{\overline\Lambda}{\pi(l+\omega')}
\right\}\nonumber\\
&-&(\omega'\rightarrow \omega)
+\frac{\eta}{{\cal R}^4}\frac{ {\overline\Lambda}^4}{2}
 \left(l-\frac{1}{2}\right)
\ln\left\{
\frac{[{\overline\Lambda}^2+\pi^2( l\pm\omega')^2]_*}
{[{\overline\Lambda}^2+\pi^2( l\pm\omega)^2]_*}\right\}
\end{eqnarray}
where the substitution $(\omega'\rightarrow\omega)$ 
only applies to terms in front of it,
to give the (separate, $\omega$ dependent) fermionic contribution.
The last term in (\ref{v2})
contains both the fermionic and the bosonic contributions.
Also
\begin{equation}
g(l,\omega')=\frac{1}{60}\left\{10-15
l+6l^2+3\left(-5\omega'+4l\omega'+2\omega'^2\right)\right\}
\end{equation}
If we sum over a finite tower of KK states (finite $l$)
we find quadratic and logarithmic 
contributions not  canceled in the sum $\Vc_1+\Vc_2$.  
If in $\Vc_2$ we consider the limit $l\gg{\overline\Lambda}$
or $l\ra \infty$ 
the logarithmic term in the first curly braces of (\ref{v2}), 
may be expanded in a power series  and its first term in this 
expansion cancels the  quadratic terms  ${\overline\Lambda}^2$ 
due to states up to level $l$, similar to the case of  $\Vc_1$.

To conclude, just as in the case of $\Vc_1$, 
quadratic or logarithmic divergences are absent  in the bosonic sector and
the same applies to the fermionic sector in the limits of either  
$l\gg (\Lambda \Rc)/2$  or of summing the
whole Kaluza Klein tower. In this mechanism, the presence of Kaluza-Klein 
modes of mass larger than the momentum cut-off 
 of the loop integral was essential. The presence or need for KK 
states beyond the momentum cut-off of the {\it effective} theory 
therefore requires a full string regularization/calculation 
of this model, where the inclusion of such states can be
addressed consistently. A detailed string  regularization/calculation 
will be provided in Sections \ref{stringreg} and \ref{stringenergy}.

\subsubsection{The minimum condition}\label{minimumcond}
We impose the minimum condition for the scalar potential with fixed
number of Kaluza Klein states (represented by $l$),  
to find the compactification radius
in terms of its value ${R_*}$ derived from the minimum condition for 
$\Vc_0$ which is the limiting case when $l\ra \infty$.
The condition for the minimum of the potential $\Vc$ leads to 
\begin{equation}\label{compact}
{\cal R}^4\approx { R_*}^4\,\, 
\frac{\sin(\pi m_t(\phi) {\cal R})}{\sin(\pi m_t(\phi) {R_*})}
\left[1-\frac{1}{6 \pi} \frac{{\cal R}^4}{\eta}
\frac{\partial (\Vc_1+\Vc_2)}{\partial \omega} \right]
\end{equation}
where ${R_*}\approx 1/(2 m_t)$ ($\sin (\pi m_t {R_*})
\approx 1$) and where  an explicit dependence of the top mass 
on the v.e.v. of Higgs  is used\footnote{This
dependence is that of \cite{Barbieri} appropriate for our analysis,
with $m_t(\phi)=2/\pi \Rc \arctan(\pi \Rc \phi y_t/2)$, but other
cases may be considered too.}. Tedious calculations give the following
approximation for the  derivative in eq.(\ref{compact}) 
\begin{eqnarray}\label{min}
\frac{{\cal R}^4}{\eta} \frac{\partial (\Vc_1+\Vc_2)}{\partial \omega}
&\approx &\frac{\pi^2}{2}
\{ 4{\overline\Lambda}^2 \omega' (l+1)
-[\pi^2
(l+\omega')^2 (l+\omega'+1)^2 \ln 
\left(1+ \frac{ {\overline\Lambda}^2 }{\pi^2 (l+\omega')^2 }\right)\nonumber\\
&-&\lf.\lf. (\omega'\ra -\omega')\right]\right\}+\frac{{\overline\Lambda}^4}{2}\ln
\frac{ {\overline\Lambda}^2+\pi^2(l+\omega')^2}
{{\overline\Lambda}^2+\pi^2(l-\omega')^2}-(\omega'\ra \omega)
\end{eqnarray}
The last substitution $(\omega'\ra \omega)$ applies to all
terms in front of it, including those within curly braces.
In the limit $l \ra \infty$ (fixed ${\overline\Lambda}$)
 the derivative of $\Vc_1+\Vc_2$ vanishes (separately for $\omega$ and 
$\omega'$ dependent terms respectively) and
we recover the case ${\cal R}={ R_*}$. We now assume there is a correlation
between the momentum cut-off of the loop integrals (\ref{pot0}),
(\ref{pot1}), (\ref{pot2}) and the number of Kaluza Klein states in 
the tower. Therefore we assume the following correlation 
\begin{eqnarray}\label{corr}
{\overline\Lambda} \approx \pi \xi l
\end{eqnarray}
which would be expected in 4D effective theory 
since $\Lambda {\cal R}$ approximates the number of
KK states between the scales $1/{\cal R}$ and $\Lambda$.
Using (\ref{corr}) in (\ref{min}) and taking again the limit $l \ra \infty$ 
of ${\partial (\Vc_1+\Vc_2)}/{\partial \omega}$ we may obtain a finite result
only if $\xi\ra 0$. This  means that no correlation between the momentum
cut-off $\Lambda$ and the (largest) mass in the KK tower (controlled
by $l$)  can be established, if the minimum condition is to be
maintained. The existence of the latter is thus a direct consequence of 
infinitely many KK states contributions and its origin should eventually 
be investigated and traced back in the context of the original 
five dimensional  theory after supersymmetry breaking.

\vspace{0.5cm}

\subsect{The scalar potential: further analysis}\label{spfurtheranalysis}
In this section we provide a different mathematical approach to the
behaviour of the scalar potential. The analysis is
closer to a string approach. In this case, while summing up infinitely
many KK modes,  the presence of KK states of
mass larger than $\Lambda$ is not manifest. 
For Case 1 below we  again  show  that quartic and
quintic divergences are canceled due to equal number of bosons and
fermions (due to initial, now broken supersymmetry), and there are
no quadratic and logarithmic divergences  in either  the bosonic or the
fermionic sectors alone, if we sum over the whole KK tower. 
This result is in agreement with our findings so far\footnote{see 
also Appendix C.} and with those of \cite{dmg}, \cite{steph}.

To compute the KK states' contributions to the potential 
one may alternatively use an integral representation of their
individual logarithmic contributions.   
This approach generalizes that using the
Riemann representation of zeta function in Appendix C (computed for 
$\omega=0$ only). Consider an individual mode $m_n$  and its contribution 
$\Vc_n$ to the scalar potential, either bosonic 
($\Vc_n^b\equiv \Vc_n(m_n\ra m_{B_n}$)) or 
fermionic ($\Vc_n^f\equiv \Vc_n(m_n\ra m_{F_n}$)) part. 
This contribution has the structure\footnote{The limits below are
taken in this order: first $x\ra 0$, then $\Lambda\ra \infty$.}
\begin{eqnarray}\label{vnn}
\Vc_n&\equiv&\frac{1}{(2\pi)^4}\int d^4 k\, \ln(k^2+m_n^2)/\Lambda^2=
\lim_{\epsilon\ra 0}\frac{2\pi^2}{(2\pi)^4}\int_{0}^{\Lambda} dk \, 
k^3 \int_{\epsilon}^{\infty}  \frac{dt}{t}\, 
\left[e^{-t }-e^{-t(k^2+m_n^2)/\Lambda^2}\right] \\
&=&\lim_{x\ra 0}
\frac{2\pi^2}{(2\pi)^4}\left\{
\frac{\Lambda^4}{4}\int_{x^2}^{\infty}\frac{dt}{t}e^{-t \Lambda^2}
-\frac{1}{2}\int_{x^2}^{\infty}
\frac{dt}{t^3} \left[e^{-t m_n^2}-e^{-t(\Lambda^2+m_n^2)}\right]
+\frac{\Lambda^2}{2}\int_{x^2}^{\infty}
\frac{dt}{t^2}e^{-t(m_n^2+\Lambda^2)}\right\}\nonumber
\end{eqnarray}
where we defined $x^2=\epsilon\Lambda^{-2}$, i.e. $x$ has dimension
(mass)$^{-1}$.
The first term in the curly braces is a quartic divergence which
cancels between bosonic and fermionic degrees of
freedom. The  second term gives  quadratic $\Lambda^2$ 
(also logarithmic $\ln\Lambda$) dependence
which can be seen by taking the limit $t\ra 0$; these terms  are the
analogue of  $m_n^2 \Lambda^2$ and $\ln(1+\Lambda^2/m_k^2)$ 
in eq.(\ref{Vn}) which may
be shown by explicitly performing the integrals in (\ref{vnn})
and then taking the limit $x\ra 0$. 
As long as one only performs a finite summation over the Kaluza Klein
level $n$ in eq.(\ref{vnn}) to obtain the scalar potential, 
the quadratic and logarithmic terms 
survive in the final result. If we perform an  infinite
summation over $n$  the  (ultraviolet) behaviour 
in $t \ra 0$ of the integrand  may change. 

From eq.(\ref{vnn}) we  find the following expression
for the  scalar potential  (in which case 
$m_n\ra m_{B_n}$ or $m_n\ra m_{F_n}$ respectively)
\footnote{We use the definition
\begin{equation}
\Gamma(\alpha,x) =\int_{x}^{\infty}dt\, e^{-t} \, t^{\alpha-1}
\end{equation}}
\begin{equation}\label{calv}
\Vc=\Vc^b-\Vc^f\equiv \frac{1}{2}(4N_c)\sum_{n} \left(\Vc_n^b-\Vc^f_n\right)=
-\lim_{x\ra 0} \frac{2\pi^2}{(2\pi)^4}\frac{1}{2}(4N_c)
\left\{\Omega\vert_{m_n\ra m_{B_n}}-\Omega\vert_{m_n\ra m_{F_n}}\right\}
\end{equation}
where only a $\Lambda^4$ term was 
canceled\footnote{the first term in eq.(\ref{vnn}).}
between the bosons and the fermions, while the remaining bosonic 
and fermionic contributions have each the structure
\begin{equation}\label{omega}
\Omega=\frac{1}{2}
\sum_{n}\int_{x^2}^{\infty} \frac{dt}{t^3}
\left\{ 1-e^{-t \Lambda^2}-t \Lambda^2 \, e^{-t \Lambda^2}
\right\} e^{-t m_n^2}
\end{equation}
We  consider the case 
when the sum  over $n$ in (\ref{calv}), (\ref{omega}) is not
restricted and is indeed performed  over all positive/negative
integers and address the implications of this procedure.
In the following we  consider for the mass $m_n$
of the $n^{th}$ KK mode  two generic  cases:

\subsubsection{Case 1.}\label{case1}
The mass is in this case assumed to be  
\begin{equation}\label{mn}
m_n=\frac{2}{{\cal R}}(n+\sigma)
\end{equation}
where $\sigma=\omega$ (fermions) or $\sigma=\omega'$ (bosons) which 
corresponds to the case of \cite{Barbieri} also discussed in 
previous sections. In the limit of including all KK states of the tower,
as can be noticed from the Kaluza Klein bosonic/fermionic 
integrals (\ref{oneint}) and (\ref{twoint}) 
of Appendix C (with $\sigma=0,1/2$), the infinite summation 
behaves like $x^{-5}=(mass)^5$ and no logarithmic or quadratic divergences
are present. The absence of the latter two 
is ensured not by supersymmetry, but by the infinite 
summation we perform, as we show below. For this consider 
a slightly different (but equivalent) form of the  
Poisson re-summed expression of (\ref{omega}) which gives that
\begin{eqnarray}\label{omegasum}
\Omega&=&
\frac{1}{2}\int_{x^2}^{\infty} \frac{dt}{t^3}
\left( 1-e^{-t \Lambda^2}-t \Lambda^2\,e^{-t \Lambda^2} \right)
\sum_{n\in\IZ} e^{-t (n+\sigma)^2/({\cal R}/2)^2}\nonumber\\
&=&\frac{\sqrt{\pi}}{2}\frac{{\cal R}}{2}\int_{x^2}^{\infty} \frac{dt}{t^{7/2}}
\left( 1-e^{-t \Lambda^2}-t \Lambda^2 \, e^{-t \Lambda^2}\right)
\sum_{k\in\IZ} 
e^{-\pi^2 ({\cal R}/2)^2 k^2/t} e^{2 i \pi k\sigma}\nonumber\\
&\equiv & \Omega_{k=0}(\Lambda)+\Omega_{k\not=0}(\Lambda)
\end{eqnarray}
where in the last step 
we separated the  contribution due to the resummed index $k=0$ (divergent)
from $\Omega_{k\not=0}$ part.
After some algebra we find
\begin{equation}\label{omega0}
\Omega_{k=0}(\Lambda)=
\frac{\sqrt{\pi}}{2}\frac{{\cal R}}{2} \frac{2}{5}
\left\{ 2 {\sqrt\pi} \Lambda^5 \left(Erf(\Lambda x)-1\right)
+{ x^{-5}}\left(1-e^{-\Lambda^2 x^2}(1+\Lambda^2 x^2-2
  \Lambda^4 x^4)\right) \right\}
\end{equation}
which in the limit of $x\ra 0$ has the following behaviour
\begin{equation}\label{limit}
\Omega_{k=0}(\Lambda)\propto
\frac{\sqrt\pi}{2}\frac{{\cal R}}{2}\frac{2}{5}\left\{
{-2 \sqrt\pi }\Lambda^5+ \frac{5}{2}\Lambda^4 x^{-1} \right\}
+{\cal O}(x)
\end{equation}
Thus all divergences ($x^{-1}\Lambda^4$ and $\Lambda^5$)
are  $\sigma$ independent,
and no quadratic or logarithmic terms are manifest anymore. 
To explain the absence of
the latter two, note that even though every integrand in the first 
line in the  r.h.s. of (\ref{omegasum}) 
has a quadratic divergence, because we sum an infinite number 
of such contributions,
the behaviour of the integral in $t=0$ changes. This may be seen after
a Poisson resummation needed to compute the leading behaviour
in $t=0$. Indeed, Poisson resummation 
transforms individual, infinitely  many  exponential contributions
($exp(-t)\ra 1, t\ra 0$)  in the r.h.s. of (\ref{omegasum}) to  
suppressed ones, $(exp(-1/t)\ra 0, t\ra 0)$.
Note also that in (\ref{omega0}) one could exchange the
order of the limits in $x$ and $\Lambda$ when evaluating
the supertrace over bosonic and fermionic contributions, since its
divergent terms cancel  anyway due to the equal number of bosonic and
fermionic degrees of freedom 
(due to the initial,  underlying supersymmetry).

Further, the contribution  $\Omega_{k\not=0}(\Lambda)$ can be integrated
from $x\ra 0$ to give\footnote{We use the notations
\begin{equation}
{\Li}_n(x)=\sum_{k=1}^{\infty}  \frac{x^k}{k^n};\,\,\,\,\,\,\,\,\,\,
\gamma(1+m,x)=m!\left\{1-e^{-x} \sum_{p=0}^{m}\frac{x^p}{p!}\right\},
\,\,\,m=0,1,2,\cdots
\end{equation}}
\begin{eqnarray}\label{omeganot0}
\Omega_{k \not=0}(\Lambda)&=&
\frac{1}{8\pi^4 ({\cal R}/2)^4} \left\{ 3 \Li_5(e^{2 i\pi \sigma})
-3 \Li_5(y)-4 {\overline\Lambda}^3 \Li_2(y)-6 {\overline\Lambda}^2 \Li_3(y)
-6 {\overline\Lambda} \Li_4(y)\right\}\nonumber\\
&&\,\,\,\,\,\,\,\,\,\,\,\,\,\,\,\,\,\,\,\,\,\,\,\,\,\,\,\,\,\,
+(\sigma\ra -\sigma)\nonumber\\
&=&
\frac{1}{8 \pi^4 ({\cal R}/2)^4} 
\left\{\frac{1}{2}
\sum_{n=1}^{\infty} 
\frac{e^{2 i \pi n \sigma}}{n^5}\,
\gamma(4, 2 n {\overline\Lambda} )+ (\sigma\ra -\sigma)
\right\}
\end{eqnarray}
where 
\begin{equation}
y=exp(-2 {\overline\Lambda}+2 i \pi
\sigma),\,\,\,\,\,\,{\overline\Lambda}={\pi\Lambda}\frac{{\cal R}}{2}
\end{equation}
Therefore  $\Omega_{k\not=0}$ is finite in the limit of large
$\Lambda$ and all  divergences of $\Omega$, of the type
$x^{-1}\Lambda^4$ and $\Lambda^5$
are present in $\Omega_{k=0}$. 
No quadratic or logarithmic  divergences are present in 
the scalar potential eq.(\ref{calv}), which supports our previous findings
where we used a  summation over an arbitrary number ($l$) of  
KK states and then took the limit $l \ra \infty$. 
It is  the  infinite number of states that is
responsible for the absence of quadratic/logarithmic
terms in either the bosonic or fermionic sectors alone \cite{dmg}. 
The remaining divergences $x^{-1}\Lambda^4$ and $\Lambda^5$
have  the same coefficient for bosonic and fermionic sectors. 
It is thus sufficient that 5D theory had equal number of bosonic and
fermionic degrees of freedom 
for these to disappear too from the scalar potential. 
Thus it is not necessary that supersymmetry be present for these
cancellations to take place. Finally, we  find from
(\ref{calv}), (\ref{omegasum}), (\ref{omeganot0}) that
\begin{eqnarray}\label{vfinale}
\Vc(\phi)&=& 
\,\,-\frac{4N_c}{2}\frac{2\pi^2}{(2\pi)^4}
\left( \Omega\vert_{\sigma\ra \omega'}-\Omega\vert_{\sigma\ra \omega}
\right)
=
\frac{3}{2 \pi^6}\frac{4 N_c}{  {\cal R}^4}
\frac{1}{3!}\sum_{k=0}^{\infty}
\frac{\cos(2(2k+1)\pi \omega(\phi))}{(2 k+1)^5} \, \gamma(4, 2 n
{\overline\Lambda} )\nonumber\\
&{\rightarrow}&
\frac{3}{2 \pi^6}\frac{4 N_c}{  {\cal R}^4}
\sum_{k=0}^{\infty}\frac{\cos(2(2k+1)\pi \omega(\phi))}
{(2 k+1)^5}
\end{eqnarray}
In the last step the limit ${\overline\Lambda}\equiv 
\pi\Lambda \Rc/2 \ra \infty$ was 
taken to recover the result $\Vc_0$ of  (\ref{el}) and already
investigated elsewhere \cite{Barbieri}. 
This means either $\Lambda$ or
$\Rc\ra\infty$ with the other fixed. 
Such limits seem unclear from the string
theory point of view we adopt.  In a string embedding of this
model, one would replace the cut-off $\Lambda$
by the string scale  $M_{string}$. Taking $\Lambda\ra\infty$
cannot be considered consistent  without turning on winding
modes, not included here. Taking instead $\Rc\ra \infty$ 
is again problematic in a string orbifold
construction as we discuss in section \ref{remarks}. 
Both limits asymptotically decouple the effects 
of winding modes, thus correspond to a string calculation in a 
very particular/singular point in the  moduli space, $T\equiv
\Rc^2/\alpha'\ra \infty$.  A full (heterotic) 
string calculation which will recover the result (\ref{vfinale}) 
as a particular case is provided in Section \ref{stringenergy}. 
However the string result cannot be applied to values of $1/\Rc$
in the TeV region\footnote{as models corresponding to Case 1 require for
phenomenological purposes.} due to perturbativity constraints on the 
10D (heterotic) string coupling which forbid large volume 
compactification.

\subsubsection{Case 2.}\label{case2}
The mass of the KK states is 
\begin{equation}\label{mnp}
m_n^2=\left[\frac{2}{{\cal R}}\right]^2(n+q)^2+M^2_\phi
\end{equation}
where  $q=1/2\,(0)$ for bosons and $q=0\,(1/2)$ for fermions respectively. 
The exact choice for charge assignment\footnote{For an (string theory)
insight into this assignement see section 3, footnote 36.}
to fermions/bosons
is crucial for the presence/absence of the electroweak symmetry 
breaking. The mass dependence (\ref{mnp})  corresponds to  models of
type \cite{antoniadis,Delgado} and bears some similarities to
heterotic  string models in the sense that the (mass)$^2$ of KK modes
is shifted by $N=M^2_\phi$ where $N$ is an excited
heavy mode of the string (for this see eq.(\ref{polylog})).

The assumption of summing over  an infinite number of KK states allows
one to perform a Poisson re-summation to find 
\begin{eqnarray}\label{omegasum2}
\Omega&=&
\frac{1}{2}\int_{x^2}^{\infty} \frac{dt}{t^3}
\left( 1-e^{-t \Lambda^2}-t \Lambda^2\,  e^{-t \Lambda^2}
\right)e^{-t M_{\phi2}^2}
\sum_{n\in\IZ} e^{-t (n+q)^2/({\cal R}/2)^2 }\nonumber\\
&=&\frac{\sqrt{\pi}}{2}\frac{{\cal R}}{2}\int_{x^2}^{\infty} \frac{dt}{t^{7/2}}
\left( 1-e^{-t \Lambda^2}-t \Lambda^2 \, e^{-t \Lambda^2}
\right) e^{-t M^2_{\phi}}\, 
\sum_{k\in\IZ} e^{-\pi^2 ({\cal R}/2)^2 k^2/t} e^{2 i \pi k q}\nonumber\\
&\equiv & \Omega_{k=0}(\Lambda)+\Omega_{k\not=0}(\Lambda)
\end{eqnarray}
where in the last step  we again separated the  contribution 
due to the re-summed KK index $k=0$ (divergent) from the 
$\Omega_{k\not=0}$ part. It is then clear that $\Omega_{k=0}$ 
part is independent of $q$ which distinguishes between bosons 
and fermions. Its 
divergent behaviour is canceled between bosons and fermions
as a consequence of equal number of bosonic and fermionic 
degrees of freedom  {\it and}, equally important,
of the summation over infinitely many KK states. Also
\begin{equation}
\Omega_{k\not=0}(\Lambda)=\Omega_1(z) -
\Omega_2(\tilde z)-\Omega_3(\tilde z)
\end{equation}
where
\begin{eqnarray}
\Omega_1 (z)&=&\frac{1}{8 \pi^4 (\Rc/2)^4}
\left\{3 \Li_5(z) +6 {\overline M_{\phi}} \Li_4(z)  
+ 4{\overline M_{\phi}}^2  \Li_3(z)+ (q\ra -q) \right\}
\nonumber\\
\Omega_2(\tilde z)&=&\Omega_1(z\ra \tilde z)\vert_{M_{\phi}
\ra (M_\phi^2+\Lambda^2)^{1/2} } \ra 0 \,\,\,\,\,if\,\,\,\,\,
\Lambda\ra\infty\\
\Omega_3(\tilde z)&=&
\frac{\Lambda^2}{4 \pi^2 (\Rc/2)^2}
\left\{ \Li_3(\tilde z)+2 \pi (\Rc/2) (M^2_{\phi}+\Lambda^2)^{1/2}
\Li_2(\tilde z)+(q\ra -q)\right\}\ra 0 \,\,\,\,if\,\,\,\,
\Lambda\ra\infty\nonumber
\end{eqnarray}
and with 
\begin{equation}{\overline M}_{\phi}=\pi M_{\phi}\frac{\Rc}{2}, \,\,\,\,\,
z=exp(-2 {\overline M}_{\phi}+2 i \pi q),\,\,\,\,\,\tilde z=
 exp(-2 \pi (\Rc/2) (M^2_{\phi}+\Lambda^2)^{\frac{1}{2}}
+2 \pi i q)
\end{equation}
In the case $M_{\phi}\ra 0$ ($\Lambda$ fix) we can 
recover $\Omega$ and 
the potential of the previous case, eq.(\ref{vfinale}). 
As in the previous case models with 
``KK regularization'' take the limit $\Lambda\ra \infty$
which at string embedding level most probably ``avoids'' any winding
modes' contribution.  Taking this limit we find 
\begin{equation}\label{lasteq}
\Omega_{k\not= 0}(\Lambda\ra \infty)=\Omega_1(z)
\end{equation}
which gives for the scalar potential
(using  (\ref{calv}), (\ref{omegasum2}), (\ref{lasteq}))
\begin{eqnarray}\label{aa2}
\Vc(\phi)&=&
\,\,-\frac{4N_c}{2}\frac{2\pi^2}{(2\pi)^4}
\left( \Omega_1\vert_{q\ra 1/2 }-\Omega_1\vert_{q\ra 0}
\right)\nonumber\\
&=&-\frac{N_c}{2 \pi^6 \Rc^4} \left\{\left[
3 \Li_5(z) +6 {\overline M_{\phi}} \Li_4(z)  
+ 4{\overline M_{\phi}}^2  \Li_3(z)+h.c.\right]_{q=1/2}-(q\ra 0)\right\}
\end{eqnarray}
result computed in a different approach in  reference \cite{Delgado}.
For $M_\phi=0$ we recover Case 1, eq.(\ref{vfinale})
with $\omega=0$. Both cases are particular results of the string
calculation, as discussed later in the text\footnote{see 
again eq.(\ref{polylog}) and text thereafter.}.

While summing over an infinite number of KK states
in (\ref{omega}) as done in Cases 1,2 above one could notice that
we interchanged the sum and the integral in eqs.(\ref{omegasum})
(\ref{omegasum2}).
This is possible if the integrand in (\ref{omega})
is exponentially suppressed in $x\ra 0$. Since this is not the case
one should introduce $(1- exp(-\rho t)),\, \rho\ra \infty$ regulator
and see if any $\rho$ presence is manifest in the final result. 
Such a regulator already exists $(1-exp(-t \Lambda^2))$ for part of
the integrand, provided by the 4D cut-off and the would-be divergent 
dependence on it (present in $\Omega_{k\not=0}$) cancels 
in the bosonic-fermionic difference. For the last term in curly 
braces in (\ref{omega})  one can again multiply it by a regulator and
show that upon taking $\rho\ra\infty$ such dependence vanishes in the
potential, thus enabling one to interchange the sum and the 
integral\footnote{See also Section \ref{hage} for a discussion at 
string level.}.

Although not fully manifest as in section (\ref{KKlimit}), the use of 
KK states of mass larger than the field theory cut-off $\Lambda$ may 
be traced in  that we performed a Poisson re-summation (which requires 
infinitely many states of mass $l/R$, $l$ unbound) and  which necessarily
includes states of mass larger than any effective field theory
cut-off. Additionally, in (\ref{omega}) we integrated 
from the (deep ultraviolet region) $t=x^2=0$, with $\Lambda$ fixed,  
which means we probe energies  beyond any fixed cut-off $1/t>\Lambda^2$. 
However, if we insist on considering only the (effective field theory)
region $1/t \ll \Lambda^2$, we find from eqs.(\ref{calv}),(\ref{omega})
\begin{equation}\label{VEFT}
\Vc=\frac{2\pi}{(2\pi)^4}\frac{4N_c}{4}
\sum_{n}\int_{x^2}^{\infty} \frac{dt}{t^3}
\left\{ e^{-t m_{F_n}^2} -  e^{-t m_{B_n}^2}
\right\}
\end{equation}
where the sum in front of the integral must be restricted to
integers giving KK masses of values smaller than the 
cut-off\footnote{Cut-off regularization of the potential
gives the same result as  proper time
regularization, see (\ref{VEFT}).} $\Lambda$.
This equation will be useful when establishing the link of
the effective field theory (with truncated KK tower) with string 
results for the vacuum energy (Section \ref{stringreg}).

\subsect{Remarks on perturbative expansion and ``Kaluza Klein 
  regularization''}\label{remarks}
The Kaluza Klein models usually 
overlook the issue of  validity of the perturbation theory they employ
and implicitly assume to hold. The presence 
of KK states (due to the extra dimension)
which have Yukawa and gauge interactions 
is  manifest as a  divergence  of the
associated couplings, which may thus lead to divergences 
for other quantities depending on them such as the Higgs mass.

As an example, consider the threshold corrections
to the gauge couplings in the 4D theory with a tower of $N=2$ KK
states (associated with an extra dimension) 
in the loop contributing to the running coupling.
Adding up the KK  states' contributions below a cut-off $n\leq
\Lambda {\cal R}$ gives
\begin{equation}\label{linear}
\alpha^{-1}|_{n\leq \Lambda {\cal R}}
 \propto \sum_{n} \ln\frac{\Lambda}{m_n}
=\sum_{n} \ln (\Lambda {\cal R})-\sum_{n}\ln n 
=\Lambda\Rc-\frac{1}{2}\ln\Lambda\Rc+\cdots
\end{equation}
Thus a linearly divergent behaviour \cite{Taylor:1988vt}
with the cut-off emerges. 
This behaviour is further supported by heterotic string calculations
of the gauge
thresholds \cite{Dixon},\cite{dkl} 
due to N=2 sectors (that KK states are part of)
in two-torus compactifications. In this case the behaviour is 
indeed power-like (quadratic). If one dimension of the two
torus is fixed to a value equal to the string length, 
the string result is then linearly
dependent on the (string) scale, in agreement with (\ref{linear})
above for the case of one additional compact dimension. 
This linear behaviour also exists in type I models in the limit of
one large compact co-dimension \cite{ABD}.
Such a behaviour of the gauge coupling above $1/\Rc$ scale can then lead to
a non-perturbative regime, more so when $1/\Rc$ has a small value (TeV).
Thus the perturbative analysis of the model considered may become unreliable. 
This situation may eventually be avoided in some (heterotic) models with $N=2$
sub-sectors which employ 
Scherk-Schwarz mechanism (defined in \ref{string2})
for the breaking  $N=2\ra N=0$ supersymmetry.
This will be examined in Sections \ref{oneloopgauge}. 
This case preserves the possibility of having only logarithmic
string thresholds to gauge couplings. In general dependence on
additional moduli exists in string models which 
can  spoil this behaviour.

The case of Yukawa couplings obtained in (5D) models with
KK states due to the extra dimension depends on
the superpotential $W$ assumed. 
Yukawa couplings running above the scale $1/\Rc$
is induced by wavefunction renormalization (due to KK states) 
and has two contributions. A gauge
contribution (independent of $W$) which  
may be linear with the scale, thus giving a linear behaviour of 
Yukawa on the high scale. 
A second contribution to Yukawa couplings 
due to  wavefunction renormalization is controlled by
$W$ alone. This part can be quadratic with the scale if
``delta-function'' 5D ``localized''  $W$  at some fixed points of the
orbifold  is present.   In  4D language it means we sum over two
towers of KK states (of different levels, since 5D momentum not conserved) 
in the one loop correction (induced by $W$) to wavefunction
renormalization, thus  to the Yukawa coupling. In this way Yukawa coupling
receives a strong  (ultraviolet) sensitivity to the  
scale/cut-off. Examples of this
type were provided in \cite{Barbieri} and  \cite{Delgado:2001xr}.
The behaviour may eventually be changed in some (heterotic) 
string models where string 
corrections to Yukawa couplings are related to gauge couplings thresholds 
(see Section \ref{oneloopyukawa}).

In calculations,  Yukawa couplings (gauge as well)
become  non-perturbative well before
reaching  the cut-off due to the large number of KK states present in 
the beta functions.  For this reason  the loop expansion 
of the effective potential which is after all a perturbative 
calculation involving Yukawa and gauge couplings may break down 
when summing up the effects of an infinite tower of KK states. 
In particular the {\it one-loop  improved} \cite{Sher} scalar 
potential in which one-loop corrected couplings are used instead of
their tree level values, is expected to differ significantly from 
its one loop expansion, 
due to the divergent behaviour of the couplings.
Alternatively, the (soft) Higgs (mass)$^2$ derived 
from $\Vc$ is proportional at one loop to Yukawa coupling \cite{Barbieri}
which in turn has a one loop correction (two loop correction 
to the Higgs mass$^2$) large due to the running of  Yukawa  coupling 
quadratically/linearly \cite{Barbieri}, \cite{Delgado:2001xr} 
with the scale. Thus two-loop Higgs (mass)$^2$ will 
receive an induced   sensitivity to the high scale
\cite{dmg} via the one-loop Yukawa couplings. An explicit example
was provided in \cite{Delgado:2001xr} where two-loop Higgs (mass)$^2$
was shown to be linearly sensitive to the ultraviolet cut-off 
(via the top coupling),
although the conclusion  was that Higgs mass is finite. This
conclusion  seems to imply that the 
Higgs mass is (two-loop) finite if
expressed in terms of (rescaled/renormalized)
Yukawa coupling\footnote{This is the non-renormalization
theorem for the superpotential at one loop in the model considered.},
but the  latter is however  linearly
divergent. Therefore the behaviour of the Higgs mass is worse than in the 
MSSM where only logarithmic dependence to the high scale exists. 
Further high scale sensitivity may be brought in by the fact that the
models are non-renormalizable. For the model  \cite{Barbieri}  a
quadratic divergence also exists \cite{toappear} 
as anticipated in \cite{dmg}.

Perturbative  constraints on the couplings, thus on the loop 
expansion of the effective potential are not easy to avoid. One 
possibility is that the compactification
scale $1/{\cal R}$ at which KK states set in is very close to the cut-off of
the model, thus rather high, ruling out models with large
compactification radius. In a (heterotic) string embedding case, 
the same situation
requires a compactification scale close to the string scale, for the
10D string coupling must be smaller than unity. Such a
condition would  mean that effective field theory
(requiring $1/{\cal R}\ll M_{string}$) may be 
unreliable and a full string calculation is needed. 
Phenomenological constraints usually require
 $1/\Rc\approx$TeV (Higgs mass proportional to $1/\Rc^2$), so  a 
type I embedding is then necessary. Thus the  question of a mechanism for 
fixing the value of $1/\Rc$ must be addressed, since this is simply
{\it ``fixed'' ad-hoc} to the TeV region by 
 phenomenology. This is important since a (one loop) Higgs (mass)$^2$ 
proportional to $1/\Rc^2$ (where supersymmetry is broken) 
with $1/\Rc$ large just restores the problem 
of quadratic divergences in the Standard Model formulation. 

Various issues related to ``KK regularization'' exist in the literature
and recent analyses \cite{john,kim}
were performed to investigate  this procedure. 
It has been shown \cite{dmg} (reviewed in section \ref{sp})
that quadratic/logarithmic divergences are absent in  the bosonic 
(also in the  fermionic) parts of the potential, 
due to summing over  an infinite number of
states in the KK tower (``KK  regularization'' limit).
In this mechanism KK states beyond the cut-off of the model played an 
important role. These remarks prompted  the conclusion \cite{dmg} which raised 
concerns on the physical  meaning of ``KK regularization'' in an 
{effective} field theory (of fixed cut-off) approach
and stressed the need for a string understanding of this 
problem\footnote{For this see Sections \ref{stringreg} and 
\ref{stringenergy}.}.
Other works \cite{kim} investigated the ``KK regularization''
to observe that it corresponds to an extremely anisotropic
distribution of the momenta; 
more explicitly $p_4$ and  $p_5=k/{\cal R}$ are highly correlated (the sum of
their squared values should be smaller than the 5D momentum cut-off,
$M_5$).  Thus
the ``KK regularization'' of summing infinitely many KK states corresponds
to $p_4^2\ll M_5^2\equiv l/\Rc$, i.e. a very anisotropic compactification.
This is actually manifest in the models examined by  observing 
that after summing the effects of infinitely many KK states, the
contribution of the 4D momentum in the loop-integrals of the Higgs 
mass is actually constrained to  very low values, as observed 
in  \cite{Barbieri}. 

To justify an effective field theory approach with KK modes
of mass larger than the momentum cut-off, one could consider the 
inclusion of some form factors ($exp(-n^2/(\Rc\Lambda)^2$) \cite{john}
to exponentially suppress the 
couplings of these KK states i.e. to ``smoothly'' decouple 
them. It was shown  that
a priori suitable decoupling form factors (exponentially
suppressed couplings) may still give an infinite result for the 
Higgs mass, although the conclusion in  \cite{john} seems 
somewhat different.  The results  are
dependent on the decoupling procedure of  these heavy
KK modes, i.e. this procedure is not clearly defined  on field 
theory grounds and does not lead to  unique results.

A cut-off regularization of the potential  (\ref{potential1}) 
is certainly not  the best procedure, 
but  is intuitive and suitable to prove the requirement of states
of masses larger than the momentum cut-off, i.e. the need for a full
string calculation where such a situation can be consistently addressed. 
This regularization will enable us to match the (field theory limit)
of the string result to that of the {\it effective} field theory with
truncated tower of KK states.
Such  momentum cut-off regularization was previously 
used  for the scalar potential (for a review see \cite{Sher}).
Adopting a dimensional regularization for the 4D momentum integral
only is not a better procedure, because we actually need a 
simultaneous treatment
of  the ultraviolet effects due to {\it all five} dimensions. Currently 
there is no regularization scheme able to treat on {\it equal footing} 
the (momentum  integral over the) four non-compact dimensions and the 
(KK sum due to the) extra compact dimension. This should be the case
as both are sources of  ultraviolet effects and 
one would expect a ``unified''
treatment of their  effects (see  \cite{steph} with applications
in \cite{toappear} and \cite{stephan2}).

Finite temperature arguments 
are sometimes used (see for example 
\cite{antoniadis,iantoniadis}) to justify a soft ultraviolet 
behaviour of the vacuum energy (or Higgs mass) in 4D effective models
with one additional compact dimension (5D supersymmetric model 
or N=2 in 4D). 
The inverse radius of the compact dimension  corresponds to the 
temperature. Supersymmetry in the initial 5D model 
then ensures a vanishing of the potential in 
the $\Rc\ra \infty$ (or $T=0$) limit. 
Finite values of $\Rc$ would then trigger only finite corrections 
to the scalar potential, by analogy with finite temperature calculations
of the free energy, since in going from $T=0$ (finite) to $T\not=0$
only finite corrections could appear. 
It seems to us that it is hard to reconcile this mechanism 
with the phenomenological requirement
to obtain a chiral spectrum in 4D, while smoothly varying the radius
from $\Rc\ra \infty$ of the N=2 supersymmetric limit in 4D
to a broken supersymmetry phase. 
An orbifold-like compactification is 
then necessary, but then the limit $\Rc\ra\infty$ on the orbifold 
does not restore the full initial supersymmetry at the fixed points.
In other words finite temperature arguments may apply to a
circle compactification of the extra dimension, but not necessarily 
to a manifold with fixed points (orbifold compactifications).  
 
We consider that one  necessarily needs a departure from the
field theory approach to models which are 
ultimately string models and for which naive 
field theory methods may not necessarily apply.

\subsect{The need for a string regularization scheme}\label{stringregneed}
One of the  compelling arguments for 
a string regularization scheme for field theory models with
Kaluza-Klein states comes somewhat surprisingly, from the field
theory itself.
The presence in the ``KK regularization'' limit 
of  states of masses larger than the momentum cut-off  
immediately raises the  question whether this 
is a legitimate procedure in an {\it effective} field
theory approach. There are two choices. The 5D field theory 
to start with is finite and has no fundamental length/cut-off 
at all, but  this seems highly unlikely, given the absence of
a description of quantum gravity in such  case. The second choice
is that it has one and correspondingly the 4D theory
will have one as well above which the inclusion/presence of 
KK states may not be easy to justify in {\it effective} field theory
of fixed cut-off. A symmetry argument may however
be used for motivating the summation over an
infinite tower of states in field theory, 
and the nature of this symmetry and its relationship to
modular invariance  will be  briefly addressed in section 
\ref{stringreg} and carefully investigated in \cite{stephan2}.
This issue is relevant since in some cases the effect of states
of mass larger than the cut-off 
is  to ensure the absence of quadratic/logarithmic 
divergences in the bosonic and in the fermionic sector respectively, 
even in the absence of supersymmetry. This prompted us to 
investigate Kaluza Klein models  in the larger 
context of string theory.

Topologically charged excitations associated with dimensional
reduction from five to four dimensions (winding modes in string
theory)  which also have mass larger than the effective field
theory cut-off also play an important UV role. 
To conclude that the ultraviolet behaviour of a
model  with  KK states only is that found in the field
theory limit is not necessarily true,  since 
the interplay between KK and
winding states cannot necessarily be described in models with Kaluza Klein
states only because modular invariance interchanges these states.
Summing up infinitely many KK states
may actually correspond, at string level,  to a very special/singular 
point in the moduli space where winding modes are decoupled, 
but the question of whether this decoupling limit is well defined 
remains to be answered\footnote{A symmetry argument may however 
be used to avoid this situation, see \cite{stephan2}}. 
In fact winding modes are infinitely massive
(thus asymptotically decoupled) only in the limit 
${\cal R}\ra \infty$ which cannot
be considered in string theory. In fact their 
density equals $1/\Rc$ \cite{Kutasov} and only vanishes when $\Rc\ra \infty$.
Hence the effect of the winding modes must  not be neglected 
and one may expect in a string
embedding of models with additional extra dimensions, the presence of
some additional (ultraviolet and/or regularization) effects due to them.
Since the effects of winding modes cannot be dealt with in a field
theory framework, it may be appropriate to 
truncate the tower of KK states to a
finite number, at a mass scale smaller than that (of the order of the
string scale) at which winding states are turned on. 
In the next section we explore the link with string theory when one
sums over either a truncated  or the whole, infinite KK tower.

Truncating the KK tower in an  effective field theory (of fixed cut-off)
approach may  be required by
the fact that in the supersymmetric limit ${\cal R}\ra \infty$ when the
vacuum energy vanishes, we have all KK states mass degenerate. 
Since the mass of KK states is $\propto l/{\cal R}$, for the limit 
$\lim_{{\cal R}\ra \infty} {l}/{{\cal R}}$
the highest KK level $l$ must  indeed be finite
in the {\it effective} field theory approach.
As discussed one  may try to avoid truncating the KK tower by 
exponentially/smoothly  decoupling heavy KK states situated above 
the cut-off, and still 
keep infinitely many KK states in EFT approach. 
However, it seems this procedure is not uniquely defined in 
field theory, therefore a full string approach  is necessary. 

One example where string theory regularizes an effective field theory
result obtained in the presence of (truncated tower of) 
Kaluza Klein states is the
case of the running of the effective gauge couplings. Kaluza Klein states
charged under the SM gauge group affect their running to give a power law 
dependence with the scale.  The effective field theory
result is thus divergent (cut-off dependent), eq.(\ref{linear}). 
There exist however
full heterotic string calculations which compute such effects
due to (two torus) 
compactifications which preserve a N=2 supersymmetric sector
\cite{Dixon}, \cite{dkl}.
Provided that the cut-off of the effective theory is smaller 
than the string scale, one can indeed  approximate the string threshold
corrections to the gauge couplings by an effective field theory result
obtained using a truncated tower of Kaluza Klein states \cite{ggr}.
The reason why the string can regularize the (otherwise divergent) 
field theory result is
in this case  due to the geometry of the string (two torus in the
heterotic case) which  manifests itself at field theory 
level as form factors  whose effect is small for
scales smaller than the string scale. Winding modes also play
an essential role in the finite character of the string result.

We therefore need a phenomenologically viable string model
whose low energy limit is the class of Kaluza Klein 
models investigated so far. If such model existed, one could 
explicitly see the regularization of the vacuum energy by the 
string, along the lines discussed for the gauge thresholds.

\subsect{String regularization of the vacuum energy}
\label{stringreg}
Before proceeding with detailed string calculations of the vacuum
energy (denoted by ${V}$),
we  anticipate the full string result of the following sections (see  section
\ref{stringenergy}),  to show how the (heterotic) string regularizes the 
corresponding effective field theory\footnote{For a related discussion
see \cite{dienes1}.}  result denoted $\Vc$.
The string result has as starting point the expression below 
(where $D=4$ non-compact dimensions, $R$ 
the radius used in string calculations expressed in $\alpha'$
units. However, in 
this section only we show explicitly the dependence on  $\alpha'=1$.)
\begin{equation}\label{fst}
V(R)=\alpha'^{-D/2}
\int_{\Fc_2} \frac{d^2 \tau}{\tau_2^{1+D/2}} \sum_{m,n \in\IZ}^{}
(-1)^m e^{2\pi i \tau_1 mn}
e^{-\pi \tau_2 \left(\frac{m^2 \alpha'}{R^2} +\frac{n^2
      R^2}{\alpha'}\right) }C_{(1,\theta)}(q,\bar q)
\end{equation}
The structure of the function $C$ above is not important
for the qualitative discussion of this section. Essentially, 
it encodes massive string 
states with masses equal or above the string scale 
$\alpha'^{-1/2}$. $\Fc_2$ is the
extended fundamental domain of integration.
The effect  of Kaluza Klein states (labeled by $m$) alone on the 
vacuum energy is obtained by discarding the 
winding modes,
\begin{equation}\label{kkonly}
V(R)_{KK\,\,only}=\alpha'^{-D/2} 
\int_{\Fc} \frac{d^2 \tau}{\tau_2^{1+D/2}} \sum_{k \in\IZ}
\left\{ e^{-\pi \tau_2 \frac{k^2 \alpha'}{(R/2)^2}}
- e^{-\pi \tau_2 \frac{(k+1/2)^2 \alpha'}{(R/2)^2}} \right\}
C_{(1,\theta)}
(q,\bar q)\ .
\end{equation}
This relation has some similarities with the (D=4) effective field theory
(EFT)  result, eqs.(\ref{calv}), (\ref{omega}) where
the limit $\Lambda\ra \infty$ is formally taken
\begin{equation}\label{VEFT1}
\Vc(\Rc)=\frac{2\pi}{(2\pi)^4}\frac{4N_c}{4}
\sum_{n}\int_{x^2}^{\infty} \frac{dt}{t^3}
\left\{ e^{-t m_{F_n}^2} -  e^{-t m_{B_n}^2}
\right\}
\end{equation}
with the replacements
$m_{F_n}^2=n^2/(\Rc/2)^2$ and $m_{B_n}^2=(n+1/2)^2 /(\Rc/2)^2$
and the sum extended over the whole KK tower.
However the integral (\ref{kkonly})
is not over the region $(x^2,\infty)$ ($x\ra 0$) 
(as in field theory (\ref{VEFT1})) 
which would make it diverge in  $t=0$, but over the extended fundamental 
domain ${\Fc}$. Given a different integration region in (\ref{kkonly})
 from the EFT case, one 
should expect a different string theory value for the KK effects alone
compared to that of EFT in the limit of summing over 
{\it all} KK states in the tower (``KK regularization''). The reason is that
the (infinite) KK tower alone integrated over $\Fc$ cannot be equal to 
the same tower integrated over $(0,\infty)$ only (to give the EFT result)
because it is the sum 
over  winding modes which  plays an essential role in
changing the integration region $\Fc$ 
into  $(0,\infty)$. This is discussed below.

The full string result eq.(\ref{fst}) \footnote{for details 
see section 3, eqs.(\ref{cosmoi}),
(\ref{cosmoii}), \req{blandine} and \req{cindy}, respectively.}  
is finite in the deep ultraviolet region ($t=0$).
First a Poisson resummation over the KK
mode $m$ is performed in (\ref{fst}) (resummed KK index $l$). 
Further, a technical argument used in the next sections\footnote{
This follows from  the arguments which precede
eq.(\ref{cosmoi}).} 
``unfolds''  the fundamental domain ${\Fc}_2$ of integration in (\ref{fst})
into the half-strip during which
winding modes $n$ in (\ref{fst}) play a  ``regularization''
role. The ``original'' winding number  is written as $n=p c$, while the
re-summed KK number $l$ is written as $l=p d$, with $(c,d)=1$ (prime integers).
The sums over $n$ and  $l$ are thus replaced by a sum over 
the (prime integer) pairs $(c,d)$ and a sum over $p$. 
The former sum  transforms $\Fc_2$ into
the half-strip to give $\tau_2\in (0,\infty)$ as in
(regularized) EFT. We are thus left with a sum over
the integer $p$. Winding modes $n\not= 0$ (also re-summed KK states, $l$) 
are manifest in the definition of the  integer $p$ that 
we sum over in the full string result. This result takes the 
following {\it generic} form 
\footnote{See again (\ref{cosmoi}); also $N$ is a heavy string mode.}
(with  model dependence only present in the coefficients $\gamma_N$)
\begin{eqnarray}\label{sf}
V(R)&=&\alpha'^{-(1+D)/2} R \int_{0}^{\infty} 
\frac{d\,t}{t^{3/2+D/2}} \sum_{p>0}^{}\sum_{N\geq 0} \left\{1-(-1)^p\right\}
e^{-\frac{\pi (R/2)^2}{\alpha' t} p^2} e^{-4 \pi t N \alpha'} \gamma_N\\
&=&
4^{D/2} \int_{\epsilon^2 \alpha'}^{\infty} \frac{d \,z}{z^{1+D/2}}
\sum_{s=-\infty}^{\infty}
\sum_{N\geq 0}\left\{ e^{-\frac{\pi z}{R^2} s^2} -e^{-\frac{\pi z}{R^2} 
(s+1/2)^2}\right\}  e^{-\pi z N} \gamma_N\ .\label{sf2}
\end{eqnarray} 
This full string result thus includes the contribution of the winding modes.
In the last step we made the change $4 t\alpha' \ra z$, we
performed a Poisson re-summation over $p$ (after including the $p=0$
case) to write the finite string result
into a form close to that of effective field theory 
calculations.
Since each of the two terms in (\ref{sf2})
is divergent in $z=0$, (see Appendix C for N=0, D=4)
we must introduce an ultraviolet  cut-off $\epsilon^2$ to avoid  their
divergence  of type $\epsilon^{-5}$ which cancels between
them\footnote{At field theory level this is ensured by  equal number
  of bosonic/fermionic degrees of freedom.}.

Note that in the deep ultraviolet region $t\ra 0$ 
most  contributions are exponentially suppressed in (\ref{sf}), 
except those of small radius, $R^2/\alpha'\ll 1$ which
may still contribute. The latter may be interpreted as 
corresponding to KK masses of value
$k/R > M_{string}$ which agrees with previous findings at effective field
theory level that KK states of mass larger than 
the momentum cut-off  have significant 
contribution. Their presence  in the string framework is 
however justified, unlike the case of (effective) field theory.

The cut-off at $\epsilon^2\alpha'$ in \req{sf2} 
thus excludes\footnote{That the lower integration range of the 
$t$--integral in \req{sf} has to be changed from $t=0$ to some finite cut-off value, 
when one  wants to interpret \req{sf} as field--theory result, may also be understood 
as follows: 
In the limit $t\equiv \tau_2\ra 0$ the integrand accounts for pure winding modes: After 
applying an $S$ transformation $\tau\ra-1/\tau$ ($(c,d)=(1,0)$) on the integrand, which 
changes this integration region to the region $\tau_2\ra\infty$
the number $p$ is converted to pure windings $n=p,\ l=0$, which 
clearly should not enter in the field--theoretical limit of \req{fst} we are discussing.}
the deep ultraviolet ($\tau_2\ra 0$) momentum region, to  retain under the
integral over $z$ only the momentum range $0\leq 1/z<1/(\epsilon^2
\alpha')$.
This procedure therefore excludes from the string modes
(that we sum over) those  of deep ultraviolet momentum which
would otherwise contribute to $V(R)$. In (\ref{sf2})
these string modes are represented by a mixture of Poisson re-summed 
winding and usual Kaluza-Klein modes. This means that the summation
over the index ``s'' in (\ref{sf2}) is replaced by two sums over $k$ and $w$
(with ``k'': Kaluza-Klein number, 
``w'': ``re-summed'' winding mode number) such that
\begin{equation}
|s|\equiv|k+w R^2\alpha'^{-1}|\leq 
\frac{R\alpha'^{-1/2}}{\epsilon},\,\,\,\epsilon>0\ \ ,\ \ k\neq 0\ .
\end{equation}
Therefore the Kaluza Klein contribution alone has to respect the same
upper bound, 
\begin{equation}
|k| \leq \frac{M_{string} R}{\epsilon}<\infty,\,\,\,\epsilon>0\ .
\end{equation}
We conclude that the effect of the Kaluza-Klein states alone ($w=0$) 
may be written in the form 
\begin{equation}\label{res}
V(R)_{KK only}=4^{D/2}
\int_{\epsilon^2 \alpha'}^{\infty} \frac{d \,z}{z^{1+D/2}}\sum_{|k| \leq
M_{string} R/\epsilon}
\sum_{N\geq 0}\left\{ e^{-\frac{\pi z}{R^2} k^2} -e^{-\frac{\pi z}{R^2} 
(k+1/2)^2}\right\}  e^{-\pi z N} \gamma_N\ .
\end{equation} 
This result due to KK states only may be compared to effective field 
theory results with KK towers. According to
(\ref{res}) the right procedure in the EFT seems that one should
sum up only a finite number of Kaluza Klein states, $k<\infty$
(since $\epsilon>0$) condition imposed to avoid the deep ultraviolet
region. The result in string theory is finite and well defined in 
this region,  eq.(\ref{sf}). 
However,  only when establishing the link with field theory
is a regulator $\epsilon >0$ required, which truncates the momentum
integration as well as the summation over the KK modes alone.

Eq.(\ref{res}) may be mapped (up to an overall factor, model dependent)
onto an EFT result of eq.(\ref{VEFT}) with 
$m_n^2 \propto N+(k+q)^2/R^2$ where  $N=0$ as in section~\ref{case1}
or $N=M^2_{\phi}$ as in section~\ref{case2}. 
The limit $1/t\ll \Lambda^2 \leq \alpha'^{-1}$
of ``truncated KK'' tower of eq.(\ref{VEFT}) is here
$1/z\ll \alpha'^{-1}$ and the KK  states of  (\ref{res}) which
give significant contribution (not exponentially suppressed)
are those for which $\pi z k^2/R^2\approx {\cal O}(1)$ i.e.
$k^2\approx R^2/z\ll (RM_{string})^2$, thus $k/R\ll M_{string}$. This is  
in agreement with  eq.(\ref{VEFT})
of including only Kaluza Klein states of mass below the cut-off
of the order of $M_s$. In this way the string provides a physical
regularization of an {\it effective}
field theory result
which includes (a truncated tower of) KK states only.

From (\ref{res}) it is obvious that for field theory 
calculations in the ``KK regularization'' limit,
in which one ``removes'' the restriction 
of summing over a finite KK number,
one ``recovers'' a full string result, eqs.(\ref{sf}),(\ref{sf2}).
However, this result actually includes ``more'' than 
that of field theory does, such as  the effects of winding modes as 
well as the modular invariance symmetry which
play a crucial role in ensuring a finite string result. 
Winding modes however do not have a correspondent at field theory level.
Therefore, if the procedure of summing the whole Kaluza Klein tower 
is justified in field theory, a 
remnant of the modular invariance symmetry present in the string calculations
should be present at the field theory level and be respected/implicitly
assumed by ``KK regularization''. This then 
explains why the field theory results obtained while  
summing over infinitely many KK states are  
compatible with a {\it full} string calculation (including momentum
{\it and} winding states) and not with the string  contribution 
due to (infinitely many) Kaluza-Klein states alone.
A possible source of such a remnant symmetry of modular invariance 
may be related to the discrete shift--symmetry eq.\req{shifts} 
for zero windings: $m\ra m+p$ (which thus allows one to sum 
over the whole KK tower in field theory).
This ``shift-symmetry''\footnote{We thank 
S. Groot Nibbelink for discussions on this issue.}
will be carefully examined in \cite{stephan2} and exploited in 
\cite{steph} in building a truly consistent (5D)  field theory 
regularization scheme.  We hope this sheds some light on the 
meaning of taking the limit of
``Kaluza-Klein regularization'' from the point of view of
string theory.

The safe procedure to follow in phenomenological studies
is to apply a field theory calculation of the scalar
potential up to the compactification scale and simply 
add to this the full string result 
as a ``threshold correction''. This procedure is similar to that 
of  adding string threshold corrections for the  gauge couplings 
to the  4D field theory value of the gauge couplings,
obtained from calculations below the compactification scale.


\def\Nc{{\cal N}}

\sect{String calculation of the vacuum energy}\label{stringenergy}

In this section\footnote{Throughout the whole section the 
compactification radius $R$ is 
measured in units of $\alpha'^{1/2}$.}
we want to address the question 
of $UV$ finiteness of the one--loop cosmological constant
in string theories with broken supersymmetry.
We consider string theories in $D$ space--time dimensions ($D\leq 9$) with one
internal dimension of size $M_{comp}=1/R$. The remaining $9-D$
dimensions are supposed to be 
of string scale size $M_{s}$. 
As we will explain in section \ref{argue}  the cosmological constant
in heterotic string theories has a much better $UV$ 
behaviour\footnote{Only harmless IR divergences due to massless states 
may show up.
These effects can be regularized by different methods.}
in contrast to type $I$. This fact justifies why in this section
we mainly focus on   heterotic string vacua.

\subsect{Cosmological constant in models with N=4 $\ra$ N=0 breaking}
\label{N4toN0}
\subsubsection{Harvey model in five dimensions}

As an example 
we want to calculate the one-loop cosmological 
constant of the model in ref. \cite{harvey}. 
This is a heterotic string model in five dimensions. Four dimensions are compactified
on a torus of string size and one dimension is compactified on a circle of radius
$R$. With respect to this circle an orbifold twist is introduced, which breaks
supersymmetry from N=4 to N=0. At the same time the gauge group is broken
from $E_8\times E_8$ to a single $E_8$ of level two.
The full calculation uses a combination of 
methods developed 
in \cite{obrien} and \cite{msi}. The starting point is the
world--sheet 
string one--loop 
integral:
\be
V(R)=\int_\Fc\fc{d^2\tau}{\tau_2}\ \fc{1}{\tau_2^{5/2}}
\sum_{(h,g)}Z_{(h,g)}(q,\ov q)
=\int_{\Fc_2}\fc{d^2\tau}{\tau_2}\ \fc{1}{\tau_2^{5/2}}\ 
Z_{(1,\theta)}(q,\ov q)\ .
\label{cosmo}
\ee
The sector $(h,g) = (1,1)$ from the untwisted sector does not 
contribute, since it has N=4
supersymmetry, i.e. $Z_{(1,1)}(q,\ov q) = 0$. The region $\Fc_2$ is
the 
(extended) 
fundamental region $\{1, S, ST \}\Fc$ of the modular subgroup 
$\Gamma_0(2)$ \cite{msi}. 
The partition function in the $(1,\theta)$ sector is
\be
Z_{(1,\theta)}(q,\ov q)=\Nc^{SO(1,1)}_{(1,\theta)}(q,\ov q)\ 
\Cc_{(1,\theta)}(q,\ov q)=
\Nc^{SO(1,1)}_{(1,\theta)}(q,\ov q)\ f_{(1,\theta)}(\ov q)\ g_{(1,\theta)}(q)
\label{partition}
\ee
with the three building blocks ($q=e^{2\pi i\tau}, \ \ov q=e^{-2\pi i\ov\tau}$).
\bea
\ds{f_{(1,\theta)}(\ov q)}&=&\ds{
\fc{\theta_4(\ov q^2)^4}{\eta(\ov q)^{12}}\ 
[\theta_3^4(\ov q)-\theta_4^4(\ov q)+\theta_2^4(\ov q)]=
\sum_{N\geq 0}c(N)\ \ov q^N} \nonumber\\
&=&32(1+8\ov q+
40\ov q^2+160\ov q^3+\ldots)\nnn
\ds{ g_{(1,\theta)}(q)}&=&\ds{\fc{E_4(q^2)}{\eta(q^2)^{12}}=
\sum_{M\geq -1} d(M)\ q^M=q^{-1}+252q+
5130q^3+54760q^5+\ldots\ ,}\nnn
\ds{\Nc^{SO(1,1)}_{(1,\theta)}(q,\ov q)}&=&\ds{
\sum_{(p_L,p_R)}(-1)^m\ q^{\h|p_L|^2}\ \ov q^{\h|p_R|^2}
=\sum_{m,n\in \IZ}(-1)^m\ e^{2\pi i \tau_1mn}\ e^{-\pi
\tau_2(\fc{m^2}{R^2}+R^2n^2)}}
\label{building}
\eea
and the Narain momenta:
\be
p_{R/L}=\fc{1}{\sqrt 2}\lf(\fc{m}{R}\mp Rn\ri)\ .
\label{onenarain}
\ee
It should be kept in mind, that the functions $f_{(1,\theta)}(\ov q)$ and 
$g_{(1,\theta)}(q)$ transform automorphic under the modular subgroup
 $\Gamma_0(2)$
\cite{msi}. Poisson resummation on $m$ in $\Nc^{SO(1,1)}_{(1,\theta)}(q,\ov q)$
yields:
\be
\Nc^{SO(1,1)}_{(1,\theta)}(q,\ov q)=\fc{R}{\tau_2^{1/2}}
\sum_{n,l\in \IZ}e^{-\fc{\pi R^2}{\tau_2}
|l+\h+\tau n|^2}\ .
\label{afterpois}
\ee
We rewrite the lattice sum \req{afterpois} in the integrand of eq. 
\req{cosmo}
\be
\sum_{n,l\in \IZ}e^{-\fc{\pi R^2}{\tau_2}
|l+\h+\tau n|^2}=\h\sum_{(n,l)\neq (0,0)}\ [1-(-1)^l]\ 
e^{-\fc{\pi R^2}{4\tau_2}|l+2n\tau|^2}\ .
\label{rewrite}
\ee
Now we shall see how the orbit decomposition of \cite{obrien} can be 
applied\footnote{The convergence of this integral depends on the 
value of the radius $R$. Below the critical radius $R_H=\h \sqrt{2}$ a 
tachyon appears in the spectrum \cite{harvey}, which
spoils the convergence. We shall be more precise about this issue in 
section \ref{hage}. } to
\be
V(R)=\h R\int\limits_{\Fc_2} \fc{d^2\tau}{\tau_2}\ \fc{1}{\tau_2^3}
\sum_{(n,l)\neq(0,0)}
[1-(-1)^l]\ e^{-\fc{\pi R^2}{4\tau_2}|l+2n\tau |^2}\ \Cc_{(1,\theta)}(q,\ov q)\ .
\label{todo}
\ee
As in \cite{obrien} we define $n=pc,\  l=pd$ with $(c,d)=1$ and the 
exponential in
\req{todo} becomes now $exp({-\fc{\pi R^2}{4\tau_2}p^2|d+2c\tau |^2})$, which
can be obtained from 
$exp({-\fc{\pi R^2}{4\tau_2}p^2})$ after the modular transformation
$\lf(\begin{array}{cc} a&b/2\\ 2c&d  \end{array}\ri)\in\Gamma_0(2)$ 
(with $ad-bc=1$) acting on $\tau$. Taking all those $\Gamma_0(2)$
elements in \req{todo}, i.e. all $(c,d)=1$ 
with the above conditions, which imply $(-1)^l = (-1)^{pd} = (-1)^p$, 
since $a,d$ are always odd integers, we unfold the fundamental region 
$\Fc_2$ in \req{todo}
to the half-strip $\Hc = \{\tau |  -\h \leq \tau_1\leq \h\ ;\ 0\leq
\tau_2<\infty\}$.  Note that these manipulations are possible, since
the modular function 
$\Cc_{(1,\theta)}(q,\ov q)$ is automorphic under the modular 
subgroup $\Gamma_0(2)$ \cite{msi}.
Finally we arrive at:
\bea
\ds{V(R)}&=&\ds{\h R\int_\Hc\fc{d^2\tau}{\tau_2^4}\sum_{p>0}\ [1-(-1)^p]\ 
e^{-\fc{\pi R^2}{4\tau_2} p^2}\ \Cc_{(1,\theta)}(q,\ov q)}\nnn
&=&\ds{\h R\int\limits_0^\infty \fc{dt}{t^4} \sum_{p>0} \sum_N\ [1-(-1)^p]\ 
e^{-\fc{\pi R^2}{4t} p^2}\ e^{-4\pi t N}\ \gamma_N\ .}
\label{cosmoi}
\eea
In the last step we performed the $\tau_1$--integration
\be\label{numbers}
g(\tau_2):=\int^{1/2}_{-1/2} d\tau_1\ \Cc_{(1,\theta)}(q,\ov q)=
\sum_N\  \gamma_N\ e^{-4\pi \tau_2N}\ ,
\ee
which projects all states to equal mass levels $M=N$.  For $\tau_2\ra 0$, the function 
$g(\tau_2)$ counts the physical states (of the sector $(1,\theta)$) \cite{Kutasov}. 
We defined 
the numbers $\gamma_N = c(N)d(N)\equiv  d_F(N) - d_B(N)$, 
which correspond to the bosonic-fermionic degeneracy at the mass level $N$:
\be
\sum_{N\geq 1}\gamma_N\ \ov q^N q^N=32\ (\ 2016\ \ov q q+820800\ 
\ov q^3 q^3+93749120\ \ov q^5q^5
+\ \ldots\ )\ .
\label{deg}
\ee
For the lowest level $N=0$ the last expression in eq. \req{cosmoi} may 
be compared with \cite{Dimopoulos}
\be
\h\gamma_0\ R\int\limits_0^\infty\fc{dt}{t^4}\sum_{p>0}\ [1-(-1)^p]
\ e^{-\fc{\pi R^2}{4t}p^2}
=\fc{128}{\pi^3}\ \fc{\gamma_0}{R^5}\sum_{p>0\atop p\ odd}\fc{1}{p^6}=
\fc{2\pi^3}{15}\ \fc{\gamma_0}{R^5}\ ,
\label{first}
\ee
which looks similar to the result of \cite{Dimopoulos}. 
Since (\ref{first}) comes from string theory, the dangerous state $p=0$ does
not occur. 
However according
to (\ref{deg}), in the particular model of \cite{harvey}, 
we have $\gamma_0=0$, which means boson-fermion 
degeneracy at the massless level. All contributions to the
cosmological constant come from 
the massive sector $N\neq 0$ and thus give an exponential
contribution.  Note also, how the 
tachyonic contribution, encoded in the $q^{-1}$ power of
\req{building} is projected out in 
\req{deg}. To perform the integral \req{cosmoi} we use eq. (3.471.12) of \cite{gradshteyn}
\be
\int\limits_0^\infty \fc{dx}{x^{1-\nu}}\ e^{-x-\fc{\mu^2}{4x}}=
2\lf(\fc{\mu}{2}\ri)^\nu K_\nu (\mu)\ ,
\label{use}
\ee 
for $|arg(\mu)| < \fc{\pi}{2}$ and $Re(\mu^2)>0$ and arrive at:
\be
V(R)=\fc{128}{R^2}\sum_{p>0,odd\atop N\geq 1} 
\gamma_N\ \fc{N^{3/2}}{p^3}\ K_3\lf[2\pi Rp\sqrt N\ri]\ .
\label{final}
\ee
One may wish to further manipulate the result \req{final}. With the 
asymptotic expansion
\cite{abra}
\be
K_3(z)=\sqrt{\fc{\pi}{2 z}}\ e^{-z}\ \lf(1+\fc{35}{8}\ 
\fc{1}{z}+\fc{945}{128}\ \fc{1}{z^2}+
\fc{3465}{1024}\ \fc{1}{z^3}+\ldots\ri)
\label{Bessel}
\ee
for large arguments $|z|,\ |arg(z)|<\fc{3}{2}\pi$, eq. \req{final}
can be casted into:
\be
V(R)=\fc{64}{R^{5/2}}\ \sum_{p>0,odd \atop N\geq 1}\ 
\gamma_N\fc{N^{5/4}}{p^{7/2}}\ e^{-2\pi Rp\sqrt{N}}\ \lf(1+\fc{35}{16\pi
  R}\ \fc{1}{p\sqrt{N}}+
\fc{945}{512\pi^2R^2}\ \fc{1}{p^2N}+\ldots\ri)\ .
\ee
The first order gives precisely the $V(R)\sim R^{-5/2}\
e^{-R\sqrt{N}}$ 
behaviour of the cosmological constant, which has been determined in \cite{harvey}.

\subsubsection{Harvey model in four dimensions}
\label{string1}
To obtain a model in four dimensions with $N=4\ra N=0$ breaking we may 
compactify one more dimension on a radius of string size.
Only  few details will change in the partition function \req{building}.
Essentially, everywhere in \req{building} one has to insert the 
lattice partition function $\tau_2^{1/2}\Nc_{R=1}(q,\ov q)$
accounting for the additional compactified dimension. 
Here, 
\be\label{string}
\Nc_{R=1}(q,\ov q)=\theta_3(q^2)\theta_3(\ov q^2)+\theta_2(q^2)\theta_2(\ov q^2)
\ee
is the partition function for the KK momenta and windings w.r.t. to a string--size ($R=1$) 
circle compactification. Thus we have:
\be
V(R)=\int_{\Fc_2}\fc{d^2\tau}{\tau_2}\ \fc{1}{\tau_2^{2}}\ 
\ds{\Nc^{SO(1,1)}_{(1,\theta)}(q,\ov q)}\ \Nc_{R=1}(q,\ov q)\ 
\ \fc{E_4(q^2)}{\eta(q^2)^{12}}
\ \fc{\theta_4(\ov q^2)^4}{\eta(\ov q)^{12}}\ 
[\theta_3^4(\ov q)-\theta_4^4(\ov q)+\theta_2^4(\ov q)]\ \ .
\label{cosmo1}
\ee
We can proceed as in the previous subsection to arrive at: 
\be
V(R)=\h R\int\limits_0^\infty \fc{dt}{t^{7/2}} \sum_{p>0} \sum_N\ [1-(-1)^p]\ 
e^{-\fc{\pi R^2}{4t} p^2}\ e^{-4\pi t N}\ \gamma_N\ .
\label{cosmoii}
\ee
Again, only the relative number of boson and fermions $\gamma_N$
\be
\sum_{N\geq 1}\gamma_N\ \ov q^N q^N=32\ \lf(2 + 2520\ q\ \ov q + 8096\ q^{\fc{5}{4}}\ 
\ov q^{\fc{5}{4}} +28224\ q^2\ \ov q^2+1231680\ q^3\ \ov q^3+
\ \ldots\ \ri)\ .
\label{deg1}
\ee
contributes in \req{cosmoii}.
We use \req{use} to evaluate:
\be\label{niki}
V(R)=\fc{93\zeta(5)\gamma_0}{4\pi^2R^4}+\fc{64}{R^{3/2}}\sum_{p>0,odd \atop N\geq 1} 
\gamma_N\ \fc{N^{5/4}}{p^{5/2}}\ K_{5/2}\lf[2\pi Rp\sqrt N\ri]\ .
\ee
When one uses the explicit representation for $K_{5/2}$ one can re-express
\req{niki} into a form closer  to that of field--theory results.
We refer the reader to section \ref{highlight}. 

One might be surprised, that the Harvey--model compactified to four dimensions 
has a non--vanishing $\gamma_0$ in contrast to five dimensions \req{deg}.
However, we compactify on an additional circle at the self--dual radius, where two additional 
gauge bosons become massless. Therefore, the relative number of bosons and fermions
changes by 2 as we go down from five to four dimensions.
This fact can be verified in \req{deg1}. The factor $32$ is the usual ground state degeneracy.

\subsect{Cosmological constant in models with N=1 $\ra$ N=0 breaking} 
\label{string2}

In the following we start with heterotic orbifold compactifications
with N=1 supersymmetry in four dimensions.
To break supersymmetry with a Scherk-Schwarz 
mechanism\footnote{Also known as 
coordinate dependent compactification.} one needs one unrotated 
compactified \dim, i.e. our N=1 orbifold has to possess 
N=2 sub-sectors\footnote{A large internal \dim\  
may lead to unwanted large gauge couplings. 
It was argued in \cite{Antoniadis3}, that this can be avoided in the {\em supersymmetric case}, 
when one discusses models
with vanishing N=2 beta--function coefficients $b_a^{N=2}$. 
As a side remark let us point out, that this is not the right condition to meet. 
Even for $b_a^{N=2}=0$ there are always additional universal gauge threshold 
corrections, which become huge for large extra dimensions \cite{gauge1}. 
The correct condition to avoid large threshold corrections is 
$b_a^{N=2}=12$ \cite{gauge1,gauge2}. In ref. \cite{gauge2} examples 
of N=1 orbifolds 
were presented, which have N=2 sectors and meet this condition. 
However, there may be further contributions to the gauge coupling, which go like 
$\ln(T_2), 1/T_2, 1/T_2^2,\ldots$ from the $D$--density, accounting 
for IR-effects, anomalies or higher loops. Their size is much smaller than $T_2$, 
but should also be canceled in addition and 
impose further string constraints on the models.},
whose twist leaves 
invariant one two--dimensional subplane \cite{Antoniadis3}.
This means that the breaking takes place in an N=2 subsector of
the full N=1 orbifold. 
One does not introduce mass shifts in the twisted sector.
Technically, this means that one breaks an N=2 subsector of the full orbifold
to N=0 and leaves the N=1 sector untouched \cite{Antoniadis3}.
In addition, one has the completely untwisted sector, the so--called N=4 sector,
which is broken to N=0 by the Scherk-Schwarz mechanism.

To be more explicit let us consider the $\IZ_2\times \IZ_2$ orbifold
with twists $\theta=\h(-1,-1,2)$ and $\omega=\h(-1,2,-1)$.
The N=2 sectors are ${\cal T}_\theta=\{(1,\theta),(\theta,1),(\theta,\theta)\}$, 
${\cal T}_\omega=\{(1,\omega),(\omega,1),(\omega,\omega)\}$ and 
${\cal T}_{\theta\omega}=\{(1,\theta\omega),(\theta\omega,1),(\theta\omega,\theta\omega)\}$.
We shall introduce a Scherk-Schwarz mechanism w.r.t. to one 
coordinate of the third
torus. Therefore only the N=2 sector\footnote{In the following, whenever we 
talk about N=2 sector we have in mind that special sector.} 
${\cal T}_\theta$ and the completely
untwisted sector ${\cal T}_0={(1,1)}$ are appropriate for this procedure.
As a concrete example for the \ss breaking w.r.t. to the latter
dimension 
(third torus subplane) 
we choose the following shifts on the internal charges \cite{Antoniadis3}:
\be
n\ra n\ \ ,\ \ m\ra m+p-\h n\ \ ,\ \ p\ra p-n\ .
\label{shifts}
\ee
Here $n,m$ are the winding and Kaluza--Klein quantum numbers and $p$
is the 
internal
$U(1)$ fermion charge. For more details see ref. \cite{Antoniadis3}.

Introducing a Scherk-Schwarz mechanism w.r.t. to an untwisted plane 
of an heterotic N=1 orbifold 
which has N=4, N=2 and N=1 subsectors will break these subsectors to
N=0, N=0 and N=1 subsectors, respectively.
Thus in the following two subsections we shall discuss separately the 
breaking $N=4\ra N=0$ in the N=4 sector ${\cal T}_0$
and the breaking $N=2\ra N=0$ in the N=2 sector ${\cal T}_\theta$.

\subsubsection{Cosmological constant for Scherk-Schwarz N=4 $\ra$ N=0}

The breaking from N=4 to N=0 represents one of the basic
supersymmetry breaking pattern triggered by a Scherk-Schwarz mechanism \cite{allSS}.
Let us calculate the cosmological constant for this case.
We consider the heterotic string compactified on five string size circles
and one circle of dimension $R$. The shifts \req{shifts} refer to this latter circle.
In the N=4 partition function
these shifts  imply a fermion charge dependent coupling of 
the circle partition function $\Nc^{SO(1,1)}(q,\ov q)$ 
to the fermionic sectors $\sum_{(\alpha,\beta)}(-1)^{\alpha+\beta} 
\ov\theta\lf[\alpha\atop \beta\ri]^4$, whose effect can be written:
\be
\Nc^{SO(1,1)}(q,\ov q)\sum_{(\alpha,\beta)}(-1)^{\alpha+\beta}\  
\ov\theta\lf[\alpha\atop \beta\ri]^4\stackrel{eq.\ \req{shifts}}{\longrightarrow}
\sum\limits_{(\alpha,\beta)}(-1)^{\alpha+\beta}
\ \Nc^{SO(1,1)}_{(\alpha,\beta)}(q,\ov q)\ \ov\theta\lf[\alpha\atop \beta\ri]^4, 
\label{insertion}
\ee
with \cite{Antoniadis3}
\be
\Nc^{SO(1,1)}_{(\alpha,\beta)}(q,\ov q)=\sum_{(\tilde p_L,\tilde p_R)}
(-1)^{n\beta}\ q^{\h|\tilde p_L|^2}\ \ov q^{\h|\tilde p_R|^2}\ ,
\ee
and: 
\be
\tilde p_{R/L}=\fc{1}{\sqrt 2}\lf(\ \fc{1}{R}\ [m+\h(\alpha+n)]\mp Rn\ \ri).
\label{narain}
\ee
With the lattice functions \req{rohmfunctions} defined in the appendix D 
we may express the r.h.s. of \req{insertion}
\be
\sum_{(\alpha,\beta)}(-1)^{\alpha+\beta}
\ \Nc^{SO(1,1)}_{(\alpha,\beta)}(q,\ov q)\ \ov\theta\lf[\alpha\atop \beta\ri]^4
=-\ov\theta_3^4\ (\Oc_0-\Oc_{1/2})+\ov\theta_2^4\ (\Ec_0-\Ec_{1/2})+
\ov\theta_4^4\ (\Oc_0+\Oc_{1/2})\ .
\label{kate}
\ee
Thus, the total partition function can be written\footnote{Whereas this form is convenient
for questions like modular invariance, 
it may be not obvious, that in the field--theory limit (zero windings)
integer KK momenta correspond to bosons and half--integer momenta to fermions. 
However, after explicitly working out the l.h.s. of \req{kate} and dropping the odd winding
functions $\Oc_0,\ \Oc_{1/2}$ one realizes, that
the $(NS,NS)$ sector comes with ${\cal E}_0$ while the $(R,NS)$ sector 
appearing with ${\cal E}_{1/2}$. Thus in the field--theory limit of \req{kate}, 
${\cal E}_0$ and ${\cal E}_{1/2}$ are to be 
associated with bosons and fermionic states, respectively.}
\be
Z_{{\cal T}_0}(q,\ov q)=
\tau_2^{-2}\ \Cc(q,\ov q)\lf[-\ov\theta_3^4\ (\Oc_0-\Oc_{1/2})+
\ov\theta_2^4\ (\Ec_0-\Ec_{1/2})+\ov\theta_4^4\ (\Oc_0+\Oc_{1/2})\ri]\ ,
\label{N=4cosmo}
\ee
with the modular function 
$\Cc(q,\ov q)=\ov\eta^{-12}\eta^{-24}E_4^2\ \Nc_{SO(5,5)}(q,\ov q)$.
The cosmological constant 
\be\label{blandine}
V_{N=4\ra N=0}(R)=\int_\Fc\fc{d^2\tau}{\tau_2}\ Z_{{\cal T}_0}(q,\ov q)=
\int_{\Fc_2}\fc{d^2\tau}{\tau_2}\fc{1}{\tau_2^2}\ (\Ec_0-\Ec_{1/2})\ \ov\theta_2^4\  
\Cc(q,\ov q)
\ee
may be evaluated in a  way similar to 
that of the previous subsection by noting that
$\Ec_0(R/2)-\Ec_{1/2}(R/2)=\Nc^{SO(1,1)}_{(1,\theta)}$. We arrive at:
\be
V_{N=4\ra N=0}(R)=\fc{93\zeta(5)\gamma^{{\cal T}_0}_0}{64\pi^2R^4}
+\fc{16\sqrt 2}{R^{3/2}}\sum_{p>0,odd \atop N\geq 1} 
\gamma^{{\cal T}_0}_N\ \fc{N^{5/4}}{p^{5/2}}\ K_{5/2}\lf[4\pi Rp\sqrt N\ri]\ .
\label{Nfour}
\ee
Again, only the relative number of boson and fermions $\gamma^{{\cal T}_0}_N$ contribute:
\be
\sum_{N\geq 1}\gamma^{{\cal T}_0}_N\ \ov q^N q^N=
32\ \lf(257 + 5120\,q^{\frac{1}{4}}\,\ov q^{\frac{1}{4}} + 
  40800\,q^\h\,\ \ov q^\h + 162560\,q^{\frac{3}{4}}\,\ov q^{\frac{3}{4}} + 
\ldots\ \ri)\ .
\label{deg2}
\ee
We assumed the five dimensional string--size 
lattice $SO(5,5)$ to be orthogonal, i.e.\\
$\Nc_{SO(5,5)}(q,\ov q)=[\Nc_{R=1}(q,\ov q)]^5$.
Let us point out that after imposing the shifts \req{shifts}
the $R\ra 1/R$ duality is no longer a symmetry of the theory.
This may be easily checked in expression \req{N=4cosmo}.

\subsubsection{Cosmological constant for Scherk-Schwarz N=2 $\ra$ N=0}
\label{highlight}

An N=2 sector of an N=1 heterotic toroidal orbifold represents an orbifold limit 
of a heterotic compactification on $K3\times T^2$.
This means that two of the four complex fermions 
$\ov\theta\lf[\alpha\atop\beta\ri]$ 
in \req{kate} 
are twisted by the orbifold group. Furthermore the bosonic oscillators and
gauge bosons are twisted.
Embedding the twist $\theta=\h(1,1,-2)$ in the partition function \req{N=4cosmo}
gives us the relevant partition function for the N=2 sector
${\cal T}_\theta$
\bea
\ds{Z_{{\cal T}_\theta}(q,\ov q)}&=&\ds{(\Oc_0-\Oc_{1/2})
\lf(-\lf|\fc{\theta_3^2\theta_4^2}{\theta_2^2}\ri|^2 g_{(1,\theta)}+
\lf|\fc{\theta_2^2\theta_3^2}{\theta_4^2}\ri|^2 g_{(\theta,1)}\ri)}\nnn
&+&\ds{(\Ec_0-\Ec_{1/2})
\lf(-\lf|\fc{\theta_2^2\theta_3^2}{\theta_4^2}\ri|^2 g_{(\theta,1)}-
\lf|\fc{\theta_2^2\theta_4^2}{\theta_3^2}\ri|^2 g_{(\theta,\theta)}\ri)}\nnn
&+&\ds{(\Oc_0+\Oc_{1/2})
\lf(\lf|\fc{\theta_3^2\theta_4^2}{\theta_2^2}\ri|^2 g_{(1,\theta)}+
\lf|\fc{\theta_2^2\theta_4^2}{\theta_3^2}\ri|^2 g_{(\theta,\theta)}\ri)\ ,}\nnn
\label{nice}
\eea
with the functions
\bea
\ds{g_{(1,\theta)}}&=&\ds{-\tau_2^{-2}\fc{E_4}{\eta^{18}\ov\eta^6}\ 
(\theta_3^4+\theta_4^4)\ \Nc_{R=1}(q,\ov q)}\ ,\nnn
\ds{g_{(\theta,1)}}&=&\ds{\tau_2^{-2}\fc{E_4}{\eta^{18}\ov\eta^6}\ 
(\theta_3^4+\theta_2^4)\ \Nc_{R=1}(q,\ov q)}\ ,\nnn
\ds{g_{(\theta,\theta)}}&=&\ds{\tau_2^{-2}\fc{E_4}{\eta^{18}\ov\eta^6}
\ (-\theta_4^4+\theta_2^4)\ \Nc_{R=1}(q,\ov q)}\ ,\nnn
\eea
which take into account the (untwisted) $E_7\times E_8$ gauge degrees of freedom, 
the other unrotated string--size internal dimension and the untwisted oscillators 
and zero modes.
As in the Harvey model, we may conveniently express the integral for
the cosmological constant
as an integral over the extended fundamental region $\Fc_2$, which accounts for the
three different N=2 subsectors $(1,\theta),\ (\theta,\theta)$, and $(\theta,1)$:
\bea
\ds{V_{N=2\ra N=0}(R)}&=&\ds{\int_{\Fc}\fc{d^2\tau}{\tau_2}
\ Z_{{\cal T}_\theta}(q,\ov q)}\nnn
&=&\ds{\int_{\Fc_2}\fc{d^2\tau}{\tau_2}\ \fc{1}{\tau_2^2}\ 
(\Ec_0-\Ec_{1/2})
\lf(-\lf|\fc{\theta_2^2\theta_3^2}{\theta_4^2}\ri|^2 g_{(\theta,1)}-
\lf|\fc{\theta_2^2\theta_4^2}{\theta_3^2}\ri|^2 g_{(\theta,\theta)}\ri)\ .}
\label{cindy}
\eea
By noting, $\Ec_0(R/2)-\Ec_{1/2}(R/2)=\Nc_{(1,\theta)}^{SO(1,1)}$ we
may proceed similarly to the previous sections to obtain for \req{cindy} 
\be
V_{N=2\ra N=0}(R)=\fc{93\zeta(5)\gamma^{{\cal T}_\theta}_0}{64\pi^2R^4}
+\fc{16\sqrt 2}{R^{3/2}}\sum_{p>0,odd \atop N\geq 1} 
\gamma^{{\cal T}_\theta}_N\ \fc{N^{5/4}}{p^{5/2}}\ K_{5/2}\lf[4\pi Rp\sqrt N\ri]\ ,
\label{Ntwo}
\ee
with the degeneracy coefficients:
\bea
\ds{\sum_{N\geq 0}\gamma^{{\cal T}_\theta}_N\ \ov q^N q^N}&=&\ds{
 -512\ \lf(\ 2 + 8\,q^{\frac{1}{4}}\,\ov q^{\frac{1}{4}} + 
  255\,q^\h\ \ov q^\h+
  1016\,q^{\frac{3}{4}}\,\ov q^{\frac{3}{4}} + 25032\,q\,\ov q + 
  94720\,q^{\frac{5}{4}}\,\ov q^{\frac{5}{4}}\ri. }\nnn  
&+&\ds{\lf.  844822\,q^{\frac{3}{2}}+\,\ov q^{\frac{3}{2}}
  3031480\,q^{\frac{7}{4}}\,\ov q^{\frac{7}{4}} + 
  22155392\,q^2\,\ov q^2 + \ldots\ \ri)\ .}
\label{deg3}
\eea
We see that in contrast to the Harvey model in $D=5$ we now have 
$\gamma^{{\cal T}_\theta}_0=-1024\neq 0$,
i.e. boson--fermion non--degeneracy already at the massless level.

Ultimately, we are interested in the cosmological constant
for our $\IZ_2\times \IZ_2$ orbifold model, whose N=1 SUSY is broken to N=0
by the \ss mechanism \req{shifts}. 
The cosmological constant of the untwisted sector \req{Nfour}
and that of the N=2 sector \req{Ntwo} are the only $R$--dependent contributions to the
complete cosmological constant.
Thus we may write:
\be\label{fcbayern}
V_{N=1\ra N=0}(R)=\fc{1}{4}V_{N=4\ra N=0}(R)+
\fc{1}{4}V_{N=2\ra N=0}(R)\ .
\ee
Using the expression eq. (8.451.6) of \cite{gradshteyn}
\be
K_{5/2}(z)=\sqrt\fc{\pi}{2z}\ e^{-z}\ \lf(1+\fc{3}{z}+\fc{3}{z^2}\ri)
\label{Besselhalf}
\ee
we eventually rewrite \req{Ntwo} and also \req{Nfour}
in terms of polylogarithms ($x=e^{-4\pi R}$), as done in 
field--theory calculations \cite{Delgado}.
Thus, \req{fcbayern} finally takes the form
\def\Li{{\cal L}i}
\be\ba{rcl}
\ds{V(R)}&=&\ds{\fc{93\zeta(5)\gamma_0}{64\pi^2R^4}+
\fc{4}{R^2}\sum_{N> 0}\ \gamma_N\ N \lf[\Li_3(x^{\sqrt N})+
\fc{3}{4\pi R \sqrt N}\ \Li_4(x^{\sqrt N})+\fc{3}{16\pi^2 R^2 N}\ \Li_5(x^{\sqrt N})\ri.}\\
&&\ds{\lf.\ -\Li_3(-x^{\sqrt N})-\fc{3}
{4\pi R \sqrt N}\ \Li_4(-x^{\sqrt N})-
\fc{3}{16\pi^2 R^2 N}\ \Li_5(-x^{\sqrt N})\ri]\ ,}
\label{polylog}
\ea\ee
with $\gamma_N=1/4(\gamma^{{\cal T}_0}_N+\gamma^{{\cal T}_\theta}_N)$.
It is interesting that eq.\req{polylog} has (up to an overall
model dependent factor, represented by $\gamma_N$) 
the same analytic form  as the scalar potential has
in (``KK regularized'') field theory, but the latter 
cannot explain/account for the contribution of winding modes' effects
which however includes.
For comparison, see subsections 
\ref{case1}, and \ref{case2}, eqs.(\ref{omeganot0}), (\ref{aa2}), 
which are just particular cases of the above result for 
$N=0$, $N=M^2_{\phi}$ respectively. The origin of this
similarity was discussed in section \ref{stringreg}
and is traced back to eq.(\ref{cosmoii}) which bears some
similarity in form to that of the scalar potential eqs.(\ref{omegasum}),
(\ref{omegasum2}) (with the formal limit $\Lambda\ra\infty$ 
in the expression of the potential). See \cite{Quiros} and 
references therein for further  (field--theoretical) details.

Although we have calculated the result \req{fcbayern} 
for a $\IZ_2\times \IZ_2$ orbifold
model (with Scherk--Schwarz breaking), up to trivial factors 
this result should also hold 
for other orbifolds vacua. Similar generalizations are known 
for N=2 orbifolds \cite{gauge2}.
In particular, the modular functions appearing in \req{nice} should serve as 
generic building blocks for N=2 orbifolds with Scherk--Schwarz breaking.

\subsubsection{Validity of perturbative string calculation and 
Hagedorn--transition}
\label{hage}

String theories in $D$ space--time dimensions with supersymmetry broken 
by a Scherk--Schwarz mechanism w.r.t. an internal dimension 
of radius $R$ behave like 
$D+1$--dimensional supersymmetric string theories at finite 
temperature $T=1/(2\pi R)$. 
It is notorious that in such theories a Hagedorn temperature exists, at
which various physical quantities develop singularities. See ref. \cite{ADK}
for an interesting recent account on this topic and further references.
In this section we have to ask what such a critical radius means 
for our cosmological constant
calculations in the previous sections.

We have seen that in perturbative heterotic string theories 
in $D$ space--time dimensions with one internal dimension 
of size $R$ (in string units)
and $9-D$ dimensions of string size  the cosmological constant always 
takes the generic form
\be
\label{generic}
V(R)=\int\limits_\Fc\fc{d\tau}{\tau_2^{D/2+1}}\ \lf[\ F_{(1,\theta)}
(\Ec_0-\Ec_{1/2})
+F_{(\theta,1)}(\Oc_0+\Oc_{1/2})+F_{(\theta,\theta)}(\Oc_0-\Oc_{1/2})\ \ri]
\ee
if supersymmetry is broken by some discrete $\IZ_2$--action
$\theta$. Here, the $F_{(h,g)}$ 
are functions depending on the compactification and supersymmetry
breaking details and the $\Ec$ and $\Oc$ are $R$--dependent 
partition functions encoding the information about
the KK and winding numbers (see Appendix D for further details).
One may worry whether the integrand is absolutely convergent for 
generic $R$, which is a necessary condition to exchange the order 
of integration and the summation.
Usually the partition functions are expanded as a series in $q$ and $\ov q$, which 
represents a good expansion as long as $\tau_2>1$ and the mass squared is positive.
Of course, all the lattice functions $\Ec$ and $\Oc$ share this property.
However, the functions $F_{(h,g)}$ may have negative powers in $q$ and
$\ov q$, which 
in the integrand lead to an  $IR$--singularity at $\tau_2\ra\infty$. 
In practice it proved to be very useful to rewrite \req{generic}:
\be
V(R)=\int\limits_{\Fc_2}\fc{d\tau}{\tau_2^{D/2+1}}\ F_{(1,\theta)}
(\Ec_0-\Ec_{1/2})\ .
\ee
Then, due to $F_{(\theta,1)}(\Oc_0+\Oc_{1/2})(-1/\tau)=
|\tau|^{2-D}F_{(1,\theta)}(\Ec_0-\Ec_{1/2})(\tau)$ negative $q,\ov q$--powers 
in $F_{(\theta,1)}(\tau)$ manifest themselves as $UV$--singularity $\tau_2\ra 0$ in 
the integrand $F_{(1,\theta)}(\Ec_0-\Ec_{1/2})(\tau)$.
The question is to what extent these negative powers 
may be canceled by positive
powers coming from the lattice functions $\Ec,\ \Oc$. Thus the lattice 
functions play the r\^ole of a regulator in the integrand.
As these lattice functions depend on the radius $R$, this r\^ole may be lost 
when $R$ reaches a critical value $R_H$, 
and the integrand becomes divergent.

For concreteness, let us investigate this issue for the calculations 
in \req{blandine} and
\req{cindy}. In the integral \req{blandine} the functions appear:
\bea\label{expansion}
\ds{-\theta_3^4(\ov q)\ \Cc(q,\ov q)\ (\Oc_0-\Oc_{1/2})}&=&
\ds{\lf(-\fc{8}{q}-\fc{160}{q^{1/2}}-\fc{512}{\ov q^{1/2}}-
\fc{1}{q\ov q^{1/2}}-\fc{20}{q^{3/4}\ov q^{1/4}}-\ \ldots\ \ri)\ 
(\Oc_0-\Oc_{1/2})\ ,}\nnn
\ds{\theta_4^4(\ov q)\ \Cc(q,\ov q)\ (\Oc_0+\Oc_{1/2})}&=&\ds{\lf(-\fc{8}{q}+
\fc{160}{q^{1/2}}+\fc{512}{\ov q^{1/2}}+\fc{1}{q\ov
  q^{1/2}}+\fc{20}{q^{3/4}\ov 
q^{1/4}}+
\ \ldots\ \ri)\ (\Oc_0+\Oc_{1/2})\ ,}\nnn
\ds{\theta_2^4(\ov q)\ \Cc(q,\ov q)\ (\Ec_0-\Ec_{1/2})}&=&\ds{\fc{16}{q}\ (\Ec_0-\Ec_{1/2})
+\ \ldots\ .}
\eea
The dots stand for terms with at least one positive power (in $q$ or $\ov q$).
First, one may wonder, whether the $16/q$--pole in the last function 
gives rise to a tachyon in the spectrum for {\em any} value of $R$. 
This may happen when we take the zero winding and momentum state $(m,n)=(0,0)$ from $\Ec_0$. 
However after Poisson resummation (cf. appendix D), the 
state  $(m,n)=(0,0)$ is removed
in the combination $\Ec_0-\Ec_{1/2}$. Moreover, since $m_R^2=0, m_L^2=-1$
it does not even fulfill level--matching.
There is also a physical argument, that the $16/q$--pole may not lead to a tachyon:
The $\theta_2$ refers to the Ramond sector, where tachyons never appear due to the 
vanishing zero--point energy in this sector.
Since there is no $(m,n)=(0,0)$ state in the function $\Oc_0$, the pole $-8/q$ in the first 
two functions of \req{expansion} does not give rise to a tachyon, either.
On the other hand, the negative power combinations $q^{-1}\ov q^{-1/2}$ 
may combine with the states $(\pm\h,\pm 1)$ from $\Oc_{1/2}$, 
to give rise to tachyons (with $N_L=0=N_R$), whose mass becomes zero at $R_H=1+\h\sqrt 2$,  
but\footnote{
Let us remind, that one of the main differences between gauge--bosons and tachyons
becoming massless at a critical radius is, that the mass of the gauge boson is positive
for $R\neq R_{cr.}$, whereas the tachyon mass is positive only for $R>R_H$ and becomes
negative for $R<R_H$. Of course, the latter fact is just the reason, why a particle, 
which becomes massless at $R_H$ is called a tachyon. E.g. the power $\ov q^{-1/2}$ is 
responsible for the gauge boson $(m,n)=\pm(\h,- 1)$ 
becoming massless at $R_{cr.}=\h \sqrt{2}$ (with $N_L=1,\ N_R=0$).} 
negative below this critical radius $R_H$.
Note, that a Poisson resummation removes 
this  state in the combination $\Oc_0-\Oc_{1/2}$, but
not in $\Oc_0+\Oc_{1/2}$.

In any case,
if the radius reaches the $Hagedorn$ radius $R_H$ (in string--units)
\be\label{Hagedorn}
R_H=1+\h\sqrt 2
\ee
tachyons appear in the spectrum and the convergence of the integrand in \req{blandine}
has to be questioned. In the previous calculations we assumed:
\be\label{cond}
R>R_H\ .
\ee
For that case our cosmological constant calculation managed to project out these potential 
tachyonic states.
This is the reason why we did not encounter any 
singularity in the final result.
Nonetheless, these states are part of the spectrum and become tachyonic 
as soon as $R$ is smaller than $R_H$.

The temperature $T_H=1/(2\pi R_H)$ associated to the 
radius \req{Hagedorn} is known as perturbative Hagedorn temperature
in heterotic string theory. Such a behaviour is quite 
generic in heterotic string models
with supersymmetry broken by a Scherk--Schwarz mechanism.
The functions appearing in \req{cindy} take an expansion similar to 
\req{expansion}.
Thus the above discussion may be repeated to verify the limiting 
radius \req{Hagedorn}
also for the integrand in \req{cindy}.
This behaviour is quite different from the model discussed in \cite{tom}, which 
interpolates between a heterotic string with gauge group $O(16)\times O(16)$ at large radii and 
$E_8\times E_8$ at small radii and no Hagedorn behaviour is met. 

Eq. \req{cond} may be in conflict, that we should have a small radius to keep the gauge 
couplings small. The gauge couplings increase with the radius due to threshold effects.
In effect they may turn the gauge couplings into the strong coupling regime where 
our cosmological constant calculations are clearly no longer valid as new non--perturbative
states (e.g. the NS 5--brane) may give rise to additional contributions.
However, there are models, where such growth of the gauge couplings is avoided
(cf. footnote 33).

\subsect{Cosmological constant: heterotic versus type $I$} 
\label{argue}
\def\ti{type $I$\ }

One of the main lessons of the previous section is that for heterotic string vacua
with supersymmetry broken by a Scherk--Schwarz mechanism w.r.t. an internal dimension 
of radius $R$ the string calculation always boils down to the generic form 
(cf. eq. \req{cosmoi}):
\be
\label{Generic}
V(R)=\int^\infty\limits_0\fc{dt}{t^{D/2+3/2}}\ \sum_{p>0}\ 
\Pi(p)\ \sum_N\ \gamma_N\ e^{-\fc{\pi R^2}{4t}p^2}\ e^{-4\pi t N}\ .
\ee
Here, $\Pi(p)$ is a projector. Without supersymmetry breaking $\Pi(p)=0$, in the simplest
models with supersymmetry breaking we have $\Pi(p)=\h[1-(-1)^p]$.

One of the intriguing features of expression \req{Generic} is, that it equally describes
the result of an (regulated) open string one--loop calculation in \ti with supersymmetry
broken by a Scherk--Schwarz mechanism.
E.g. the closed string result \req{blandine} has its \ti counter part 
as open descendants  
\be\label{basicrohm}
V(R)=\int\limits_0^\infty\ \fc{dt}{t^3}\ 
\theta_4(it/R^2)\ \Theta(it)\ +\ reg.\ ,
\ee
with the modular function \cite{Rohm}
\be
\Theta(it)=\lf[(d_E+d_O)\ \fc{\theta_2\lf(\fc{it}{2}\ri)^4}{\eta\lf(\fc{it}{2}\ri)^{12}}+
(d_E-d_O)\ \fc{\theta_2\lf(\fc{it}{2}+\h\ri)^4}{\eta\lf(\fc{it}{2}+\h\ri)^{12}}\ri]\ 
\theta_3(it)^5=\sum_{N\geq 0}\ 
\tilde \gamma_N \ e^{-\pi tN}\ 
\ee
encoding the open string spectrum and $\theta_4$ 
accounting for the KK states of the open string.
The two contributions to the cosmological constant originate from the annulus and M\"obius
amplitude, respectively. The Klein bottle contribution vanishes. For a recent review
and a detailed list of references about supersymmetry breaking in \ti we refer the reader to  
\cite{typeI}. The coefficients $d_E,d_O$ are Chan--Paton factors.
We shall comment on the second term, which comes from a regularization in a moment.
After a Poisson resummation in $\theta_4(it/R^2)$ we obtain
\be\label{clau}
V(R)=\h R\int\limits_0^\infty\ \fc{dt}{t^{7/2}}\ \sum_{p>0}\ 
[1-(-1)^p]\ e^{-\fc{\pi R^2}{4t}p^2}\ \sum_{N\geq 0}\ 
\tilde \gamma_N \ e^{-\pi tN}\ +\ reg.\ ,
\ee
which should be compared/contrasted with \req{Generic}.
However, there is one important issue to respect regarding the second term. 
The \ti expression \req{basicrohm} receives an $UV$ divergence\footnote{
It has been already pointed out in \cite{Rohm}, that also this \ti model has a 
critical radius $R_H$, below which string excitations $N$ become less massive than 
the momentum excitations, 
and the cosmological constant \req{basicrohm} diverges. In this case
the regularization 
method by which the integral \req{basicrohm} is defined is
insufficient to contr\^ol that 
divergence.} $t\ra 0$ for $N=0$ 
from the zero momentum state $m=0$ in $\theta_4(it/R^2)$. 
Usually this state is present in any one--loop \ti calculation and 
has to be regularized, which
gives rise to this additional term.
This is a notorious problem from field--theory with KK modes (cf. section 2).
Since an open string does not have windings, an one--loop open string 
calculation always reduces to a (naive) field--theoretical calculation 
with a tower of massive string states
(encoded in $\Theta$) and one should expect potential $UV$--problems.

It is argued in \cite{bf}, 
that the state $m=0$ is not present if \req{basicrohm} is uniformly
regularized in the transverse closed string channel and the
one--particle reducible diagrams 
are subtracted \cite{bf}. Another 
argument for the absence of the divergence is, that when 
summing over all one--loop open string diagrams the divergence drops out if certain conditions 
on the spectrum, which guarantee tadpole cancellation, hold.
In practice this means that one performs a Poisson resummation
on $\theta_4(t)$ in \req{basicrohm}, which converts the open--channel
KK momenta to the 
closed--channel windings and subtracts the contribution of 
the zero--winding state ("Poisson--resummation prescription").
This situation reminds us of that of \cite{Dimopoulos} 
and of the discussion in Section 2.

Nevertheless it is  remarkable, that the result of an one--loop heterotic string 
calculation with both KK and winding states looks like an (carefully regularized) 
open string one--loop calculation with only KK modes running in the loop. 
In particular, the appearance of windings and/or 
modular invariance tells us, that this correspondence follows to lesser extent from
the common underlying four--dimensional effective field theory\footnote{This fact just 
explains that the cosmological constant is given by some sort of 
Schwinger one--loop integral with massive states running in the loop.} 
but may be a consequence of a similar underlying regularization procedure using 
windings (and implying some sort of modular invariance) in both theories.
Open strings with non--vanishing $NS$ $B$--field, whose KK modes look like windings, 
may shed more light on this connections \cite{work}.

The analogy between the heterotic \req{Generic} and \ti result \req{clau} may sound 
familiar from heterotic \ti dualities, where one--loop heterotic
calculations are 
mapped to
\ti one--loop results. However, this duality is somewhat more subtle as
e.g. in nine (or eight)
dimensions an exact one--loop heterotic result comprises tree-- up to 
three--loop results on the \ti side \cite{hettypeI,kristin}.
Moreover, non--perturbative corrections below nine dimensions complicate this map.
But it is  remarkable and close to our discussion, that e.g. the 
leading\footnote{That part of the full heterotic one--loop
  calculation, which is mapped 
to a one--loop corrections on the \ti side. As said before, the 
remaining parts are mapped to tree-- and 
higher than one--loop parts on type $I$.} part of the 
one--loop heterotic threshold of the ${\rm Tr} F^4$ coupling in nine dimensions 
gives \req{Generic} with $\Pi(p)=1$ (and $N=0$).
Heterotic--\ti duality maps this correction, which is finite, to 
an one--loop open string 
result, written in the closed string--channel with the zero winding dropped: 
$\int\limits_0^\infty \fc{dt}{t^2}\sum\limits_{n\neq 0} e^{-2\pi R^2n^2/t}$.
In fact, the heterotic zero winding and KK contribution maps to a tree--level \ti.

\subsect{Wilson line dependence of the cosmological constant}
\label{wilsondependence}
In this subsection we want to investigate how our previous results for the cosmological constant
change under the inclusion of one Wilson line w.r.t. to the gauge group.  
The mass of the heavy particles running in the loop depends on 
this additional modulus and thus gives rise for a Wilson line dependence of the cosmological 
constant.

We introduce the (continuous) Wilson line 
\be\label{wilson}
A_i^I=a\ (\ 1,\ -1\ ,\ 0^6\ )
\ee 
(written in an $E_8$ orthonormal basis) in the second $E_8$ gauge group. 
This choice makes sure, that an $SU(2)$ subgroup of the $E_8$
is broken at generic values of $a$. In fact, \req{wilson} reads
$a_i^I=a\ (\ 1\ ;\ 0^7\ )$ w.r.t. an $SU(2)\times E_7$ root lattice basis.
With this choice \req{wilson} the Narain momenta \req{narain} change as usual
(see also \cite{karim}):
\be
\label{narainch}
\tilde p_{R/L}=\fc{1}{\sqrt 2}\lf(\ \fc{1}{R}[m+\h(\alpha+n)+\h A^2n+AN]\mp Rn\ \ri)\ .
\ee
with the $E_8$ gauge charges $N_I$.
These momenta enter the lattice functions \req{rohmfunctions}, which 
account for the different spin--structures. However, now they are also coupled 
with the gauge group: 
\be\label{newf}
\Ec_{\delta/2}=\sum_{n,m\atop N\in E_8}\ 
e^{-2\pi i \tau_1 (m+\fc{\delta}{2}+\h A^2n +N) n}\ 
e^{-\fc{\pi\tau_2}{R^2}\lf[(m+\fc{\delta}{2}+\h A^2n +AN)^2+n^2R^4)\ri]}
\ q^{\h(N+An)^2}
\ee
with $\delta=0,1$ and similarly for $\Oc_{\delta/2}$ with odd windings $n$.
After Poisson resummation on $m$ they become:
\be\label{newff}
\Ec_{\delta/2}=\fc{R}{\tau_2^{1/2}}\ \sum_{n,l\in\IZ}
\ (-1)^{l\delta}\ e^{2\pi i(\h lnA^2+lAN)}\ e^{-\fc{\pi R^2}{\tau_2}|l+\tau n|^2}
\ q^{\h(K+An)^2}\ .
\ee
Due to the special choice \req{wilson}, which breaks only an $SU(2)$ subgroup of the $E_8$ 
gauge group, it is possible to disentangle the remaining $E_7$ partition function in
the lattice sums \req{newf} and \req{newff}. As explained in appendix A of \cite{gauge2}, this 
is accomplished by splitting the $E_8$ character function into a product of 
$E_7$ and $SU(2)$ characters:
\be\label{subst}
E_4(q)=\sum_{N\in E_8} q^{\h N^2}\ra \sum_{K\ even}\ q^{\fc{1}{4}K^2} E^{even}_{4,1}(q)+
\sum_{K\ odd}\ q^{\fc{1}{4}K^2} E^{odd}_{4,1}(q)
\ee
and analogous
\be\label{split}
\sum_{N\in E_8} q^{\h (N+An)^2}\ra \sum_{K\ even}\ q^{\fc{1}{4}(K+2an)^2} E^{even}_{4,1}(q)+
\sum_{K\ odd}\ q^{\fc{1}{4}(K+2an)^2} E^{odd}_{4,1}(q)\ ,
\ee
with the $E_7$ character functions:
\bea
\ds{E_{4,1}^{even}(q)}&=&\ds{\theta_3(q^2)^7+7\theta_3(q^2)^3\theta_2(q^2)^4=1+126\ q+756\ q^2+
\ldots\ ,}\nnn
\ds{E_{4,1}^{odd}(q)}&=& \ds{\theta_2(q^2)^7+7\theta_2(q^2)^3\theta_3(q^2)^4=56\ q^{3/4}+
576\ q^{7/4}+1512\ q^{11/4}+\ldots\ .}
\eea
Thus, the lattice partition function \req{newff} can be written:
\bea\label{newfff}
\ds{\Ec_{\delta/2}}&=&\ds{\fc{R}{\tau_2^{1/2}}\ \sum_{n,l}
\ (-1)^{l\delta}\ e^{-\fc{\pi R^2}{\tau_2}|l+\tau n|^2}}\nnn
&\times&\ds{\lf\{
\sum_{K even}\ e^{2\pi i(lna^2+alK)}\ q^{\fc{1}{4}(K+2an)^2}\ E_{4,1}^{even}(q)+  
\sum_{K odd}\  e^{2\pi i(lna^2+alK)}\ q^{\fc{1}{4}(K+2an)^2}\ E_{4,1}^{odd}(q)\ri\}\ .}
\eea
With this in mind we introduce the Wilson line \req{wilson}
in the partition functions
\req{N=4cosmo} and \req{nice}: The lattice functions $\Ec_{\delta/2}$ and $\Oc_{\delta/2}$ 
have to be replaced by
\req{newfff} and one $E_4(q)$ from the gauge partition function has to be dropped because
it is already contained in \req{newfff}.
For the $\tau$--integration we can make use of the same techniques as before, i.e.
only the sector with the combination $\Ec_0-\Ec_{1/2}$ contributes at the cost of 
extending the integration from $\Fc$ to $\Fc_2$. Furthermore, the orbit decomposition
allows us to stick to the case $n=0\ ,\ l=p$ by extending the integration $\Fc_2$
to the full strip $\Hc$. Thus, we end up with integrals:
\be
\label{withWL}
\Lambda(R,a)=R\int\limits_0^\infty \fc{dt}{t^{7/2}} \sum_{p>0} \sum_{N\geq 0}\sum_{K \in \IZ}\  
[1-(-1)^p]\ 
e^{-\fc{\pi R^2}{t} p^2}\ e^{2\pi i a pK}\ e^{-4\pi t N}\ \gamma(N-\fc{1}{4}K^2,N)\ ,
\label{schalke}
\ee
which following the previous sections integrates to:
\bea\label{wlresult}
\ds{\Lambda(R,a)}&=&\ds{\fc{3}{2\pi^2R^4}\sum_{p>0, odd}\ \sum_{K=\{0,\pm 1,\pm 2\}}
\ \fc{e^{2\pi i ap K}}{p^5}\ \ \gamma(-\fc{1}{4} K^2,0)}\nnn
&+&\ds{\fc{2^{9/2}}{R^{3/2}}\sum_{p>0, odd\atop N\geq 0}\sum_{K \in \IZ} 
\ e^{2\pi i ap K}\  
\gamma(N-\fc{1}{4}K^2,N)\ \fc{N^{5/4}}{p^{5/2}}\ K_{5/2}\lf[4\pi Rp\sqrt N\ri]\ .}
\eea
The expansion coefficients $\gamma(M,N)$ are given in appendix F for the model discussed in 
section 3.3.
Again, only massless states, classified w.r.t to their $SU(2)$ charge $K$ contribute
to the first term of \req{wlresult}.
Thus essentially, a Wilson line gives rise to an additional "projector"
$e^{2\pi i a pK}$. Its effect can be easily seen for discrete values of $a$.

\subsect{One--loop corrections in heterotic Scherk--Schwarz models}
\label{sonntag}

In this section we want to investigate the general structure of one--loop
corrections in heterotic string theories in $D$ space--time dimensions 
with one large extra dimension of size $R$ and supersymmetry broken by a Scherk Schwarz 
mechanism \req{shifts}.
The analysis will enable us to give the 
explicit form for the one--loop gauge threshold corrections in such models.
From the previous subsections it is obvious, that the generic expression reads:
\be\label{Todo0}
I_{D,n}(R)=\int_{\Fc_2}\fc{d^2\tau}{\tau_2^{D/2-n+1}}\  
[\Ec_0(R)-\Ec_{1/2}(R)]\ \Cc(q,\ov q)\ .
\ee
We assume, that the condition  $D/2\geq n$ holds.
The integer $n$ depends on what kind of amplitude we are calculating.
E.g. $n=0$ for a cosmological constant calculation and $n=2$ for gauge thresholds.
Furthermore, the function $\Cc(q,\ov q)$ is a weighted partition function 
(elliptic genus) encoding the massive string states. It has modular weight
$n+1/2-D/2$ under modular transformations of both $\tau$ and $\ov\tau$.
After using the procedure described in subsection \ref{N4toN0} the integral $I_{D,n}(R)$
takes the generic Schwinger--integral form (cf. eq. \req{cosmoi}):
\be\label{Todo}
I_{D,n}(R)=R\int_0^\infty\fc{dt}{t^{D/2+3/2-n}}\sum_{N\geq 0}\ \gamma_N\ 
\sum_{p>0}\ [1-(-1)^p]\ e^{-\fc{\pi R^2p^2}{t}}\ e^{-4\pi N t}\ .
\ee
We suppose, that the analog of the $\tau_1$--integration \req{numbers} cancells any negative
$N$ powers in \req{Todo}. We point out 
that the integrand is only convergent for the case \req{cond} and the
limit $R\ra \infty$ has to be taken with care. In this limit supersymmetry is restored
and the projection onto odd numbers $p$ is suspended\footnote{For completeness, we may present
the analog of \req{Todo0} for the supersymmetric case. Then we have an unshifted 
Narain lattice for the bosonic zero modes of the internal dimensions (cf. \req{building})
and integrate only over the fundamental domain $\cal F$:
\be
I^{susy}_{D,n}(R)=\int_{\Fc}\fc{d^2\tau}{\tau_2^{D/2-n+1}}\  
\sum_{(p_L,p_R)}\ q^{\h|p_L|^2}\ \ov q^{\h|p_R|^2}\ \Cc(q,\ov q)=R\int_\Fc
\fc{d^2\tau}{\tau_2^{D/2-n+3/2}}\sum_{(n,l)}
\ e^{-\fc{\pi R^2}{\tau_2}|l+\tau n|^2}\ \Cc(q,\ov q)\ .
\ee
As usual, we performed a Poisson--resummation on the momentum $m$ (cf. \req{afterpois}).
Apart from $\Cc(q,\ov q)$, which depends on the number of supersymmetries and may be even zero
for certain couplings, 
the main difference to \req{rewrite} is, that the sum also includes the state $(n,l)=(0,0)$. 
It gives rise to a linear $R$--dependence in $I^{susy}_{D,n}(R)$ 
$$R\gamma:=R\int_\Fc \fc{d^2\tau}{\tau_2^{D/2-n+3/2}}\ \Cc(q,\ov q)\ ,$$
in contrast to $I_{D,n}(R)$. At present, we do not have a general expression for $\gamma$, 
except for $D/2-n=1/2$. In that case, $\gamma=\fc{\pi}{3}\lim\limits_{\tau_2\ra 0} g(\tau_2)$, 
provided the (in that case) modular invariant function $\Cc(q,\ov q)$ 
meets certain conditions \cite{Kutasov}. For a definition of $g$, see \req{numbers}.
Altogether, we obtain for $I^{susy}_{D,n}(R)$:
\bea\label{nyc}
\ds{I^{susy}_{D,n}(R)}&=&\ds{R\gamma+
2R\sum_{N> 0}\sum_{p>0}\lf(\fc{2N^{1/2}}{Rp}\ri)^{\h+\fc{D}{2}-n}\ \gamma_N\ 
K_{1/2+D/2-n}\lf(4\pi Rp\sqrt{N}\ri)}\nnn
&+&\ds{\hspace{-0.25cm}\lf\{\gamma_0\  \pi^{-\h-\fc{D}{2}+n}\ R^{2n-D}
\ \Gamma(1/2+D/2-n)\ \zeta(1+D-2n)\ ,\ D\neq 2n
\atop  \gamma_0\lf(\h\gamma_E-\ln R-\h\ln 4\pi\ri)\ ,\ D=2n\ .\ri.\ }
\eea}.
There will then be also
a contribution from the so--called zero--orbit $p=0$ (given by the first term of 
\req{nyc}). Thus in the following we shall assume $R\neq \infty$.

The contributions $N\neq 0$ and $N=0$ are evaluated, separately. 
For $N\neq 0$, \req{Todo} gives
\be\label{inter}
4R\sum_{N> 0}\sum_{p>0,odd}\lf(\fc{2N^{1/2}}{Rp}\ri)^{\h+\fc{D}{2}-n}\ \gamma_N\ 
K_{1/2+D/2-n}\lf(4\pi Rp\sqrt{N}\ri)
\ee
and for the $N=0$ term\footnote{This contribution
is reminiscent from an one--loop field--theory calculation with Kaluza--Klein modes 
running in the loop (cutoff sent to infinity).} we obtain
\be\label{inter1}
2R\gamma_0\int_0^\infty\fc{dt}{t^{\fc{D}{2}+\fc{3}{2}-n+\epsilon}}\sum_{p>0,odd}\ 
e^{-\fc{\pi R^2 p^2}{t}}\ .
\ee
For $N=0$ the parameter $\epsilon$ has to be introduced for the case $D/2=n$,
to regularize the integral \req{inter1}. This corresponds to some sort of 
dimensional regularization. See also Appendix C for its evaluation
using a different procedure and ref. \cite{kristin} for a similar
situation and more details. To further simplify \req{inter}, we
restrict to even space--time dimensions $D$.
In that case, the explicit representation for $K_{1/2+D/2-n}$ in
eq. (8.468) of \cite{gradshteyn} can be used to proceed further.
After some algebra eqs. \req{inter} and \req{inter1} result in
\bea\label{Todoi}
\ds{I_{D,n}(R)}
&=&\ds{\fc{2^{D/2-n}}{R^{D/2-n}}\sum_{N>0} \gamma_N\ 
N^{\fc{D}{4}-\fc{n}{2}}\sum_{k=0}^{D/2-n}\fc{(D/2-n+k)!}{4^kk!(D/2-n-k)!}\ 
\fc{1}{(2\pi R \sqrt{N})^k}\ \Li_{k+1+D/2-n}\lf(x^{\sqrt{N}}\ri)}\nnn
&-&\ds{\hspace{1cm}(x\ra -x)}\nnn
&+&\ds{\hspace{-0.25cm}\lf\{\gamma_0\  (2-2^{2n-D})\pi^{-\h-\fc{D}{2}+n}\ R^{2n-D}
\ \Gamma(1/2+D/2-n)\ \zeta(1+D-2n)\ ,\ D\neq 2n
\atop  \gamma_0\lf(\h\gamma_E-\ln R-\h\ln\pi\ri)\ ,\ D=2n\ ,\ri.\ }
\eea
with $x=e^{-4\pi R}$. 
Though we restricted to even dimensions $D$, the last term in \req{Todoi}, which comes from
the integral \req{inter1}, is valid for any integer $D$.
On the other hand, this term also shows us, that the structure of the one--loop 
threshold correction \req{Todo0} is quite 
different for even and odd dimensions $D$.
For odd dimensions the Riemann $\zeta$--function becomes elementary 
$\zeta(2s)=\sum\limits_{n=1}^\infty \fc{1}{n^{2s}}=(-1)^{s+1}\fc{(2\pi)^{2s}}{2(2s)!}B_{2s}
\ ,\ s\geq 1$, with no need for a $\zeta$--function regularization.
It is one of the reasons, why field--theories in odd dimensions have a better $UV$
behaviour.

\subsubsection{One--loop gauge couplings in theories with broken
  supersymmetry} 
\label{oneloopgauge}
Let us apply these results for the gauge--threshold corrections in four space--time 
dimensions $D=4\ ,n=2$, in particular for the model discussed in section \ref{string2}. 
We remind the reader that the model has an N=4 subsector ${\cal T}_0$ and an N=2 subsector 
${\cal T}_\theta$, which are broken to N=0 by the Scherk--Schwarz mechanism.
Furthermore an N=1 subsector, which is not affected by this mechanism. 
The former two subsectors ${\cal T}_0,\ {\cal T}_\theta$ will give rise to 
radius dependent contributions to the gauge couplings, while the N=1 sector will not.
For gauge thresholds the function
$\Cc(q,\ov q)$ has to account for the gauge and fermion charge insertion: 
${\rm Tr}\  Q_a^2 F^2 (-1)^F$ \cite{Dixon}. Technically, this means, that the 
(light--cone) space--time fermion partition function $\fc{\theta_2(\ov q)}{\eta(\ov q)}$
from the $RNS$ sector in \req{N=4cosmo} 
and \req{nice} has to be replaced by 
$\theta_2(\ov q)\ra 
\fc{1}{24}[\theta_3(\ov q)^4+\theta_4(\ov q)^4+\widehat{E_2}(\ov q)]\theta_2(\ov q)$
and similarly in the other sectors.
Furthermore, the $E_8'$ gauge partition function $E_4(q)$ has to be changed 
to $\fc{1}{3}[\widehat{E_2}(q)E_4(q)-E_6(q)]$, if
we are interested in thresholds w.r.t. to this gauge group $G_a=E_8$.
Since $\widehat{E_2}(q)=E_2(q)-\fc{3}{\pi\tau_2}$ we encounter additional
$1/\tau_2$ powers steaming from pinching the positions of the two
gauge 
vertex operators. 
They bring back the original power of $1/\tau_2^3$ of the cosmological
constant 
calculation
in \req{Todo0}, i.e. $I_{4,0}(R)=V(R)$, which has been the subject of
the preceding subsections.
In supersymmetric vacua this term is canceled by the spin structure sum \cite{Dixon}.
In total, the gauge thresholds assemble from the three contributions
\be\label{assemble}
\Delta_{a}(R)=\fc{9}{\pi^2}V(R)-\fc{3}{\pi}I_{4,1}(R)+I_{4,2}(R)
\ee
which are classified by their power in $1/\tau_2$ (besides their functions $\Cc(q,\ov q)$).
The relevant functions $\Cc(q,\ov q)$ for the model of section \ref{string2} 
will be detailed in appendix E.
The second expression summarizes all other universal corrections to the gauge couplings, 
whereas the last term $I_{4,2}$ comprises their group dependent part.
See ref. \cite{gauge1} for a detailed discussion about these terms in toroidal orbifolds.
The contribution $I_{4,2}$ has only a total $1/\tau_2$ power in the integrand
of \req{Todo0} which gives rise to the logarithmic contributions 
of the gauge threshold
corrections in four space--time dimensions. 
In that case the integrand has to be regularized, which
results in additional $\ln(R)$ terms and some scheme dependent constants
(cf. last line of eq. \req{Todoi}).

It may be a surprise, that the cosmological constant appears in \req{assemble}.
However, the following argument explains, why the latter has to appear 
in the gauge threshold result for non--supersymmetric vacua: 
If these corrections are calculated in a background gauge field
method, gravity has to be switched on to fulfill Einstein  equation.
The latter are supplemented by a cosmological constant term and 
should therefore modify the gauge threshold result. This is a novel effect 
in contrast to threshold corrections in supersymmetric string vacua 
discussed in \cite{Dixon}.

Using the ingredients from appendix E and the result \req{Todoi} the 
one--loop gauge couplings \req{assemble} for the model of section \ref{string2} take the form:
\bea\label{gaugeres}
\ds{\Delta_a(R)}&=&\ds{\fc{9}{\pi^2} \fc{93\zeta(5)\gamma_0}{64\pi^2R^4}-
\fc{3}{\pi}\fc{7\zeta(3)\beta_0}{8\pi R^2}+
\alpha_0\lf(\h\gamma_E-\ln R-\h\ln\pi\ri)}\nnn   
&+&\ds{\fc{9}{\pi^2}\fc{4}{R^2}\sum_{N>0} \gamma_N\ 
N\sum_{k=0}^{2}\fc{(2+k)!}{4^kk!(2-k)!}\ \fc{1}{(2\pi R \sqrt{N})^k}\ 
\Li_{k+3}\lf(x^{\sqrt{N}}\ri)}\nnn
&-&\ds{\fc{3}{\pi}\fc{2}{R}\sum_{N> 0}\beta_N\ 
N^{1/2}\sum_{k=0}^{1}\fc{(1+k)!}{4^kk!(1-k)!}\ \fc{1}{(2\pi R \sqrt{N})^k}\ 
\Li_{k+2}\lf(x^{\sqrt{N}}\ri)}\nnn
&+&\ds{\sum_{N>0} \alpha_N\ \Li_{1}\lf(x^{\sqrt{N}}\ri)}-\hspace{1cm}(x\ra -x)\ .\nnn
\eea
The first 
line gives the contributions from the massless sector $N=0$, whereas the
polylogarithms account for the massive string states with mass level $N$. They are 
exponentially suppressed with the radius $R$.
The second line is nothing else than \req{polylog}.
The coefficients $\alpha_N=1/4(\alpha_N^{{\cal T}_0}+\alpha_N^{{\cal T}_\theta})$
and $\beta_N=1/4(\beta_N^{{\cal T}_0}+\beta_N^{{\cal T}_\theta})$
can be picked up from the appendix, whereas $\gamma_N$ is introduced in \req{polylog}.

As we have pointed out above: In the limit $R\ra\infty$ supersymmetry is restored.
In that case the state $p=0$, i.e. $(l,n)=(0,0)$ in \req{Todo},  
gives a term linear in $R$ in agreement with gauge thresholds results for supersymmetric 
vacua with four or eight supercharges (cf. \req{nyc})  
(for sixteen and more supercharges, ${\cal C}(q,\ov q)=0$). 
One interesting observation to make at \req{gaugeres} is that the gauge thresholds do
not have a linear power dependence in $R$, which means, that the gauge couplings 
do not blow up in the decompactification limit. 
It is clear, that the N=4 sector (entering \req{gaugeres} through the coefficients 
$\alpha_N^{{\cal T}_0}$ and $\beta_N^{{\cal T}_0}$) should show such a 
behaviour, because N=4 supersymmetry is restored in the decompactification limit.
This behaviour has also been demonstrated for heterotic models with $N=4\ra N=2$
Scherk--Schwarz supersymmetry breaking \cite{KKPR}. However, we point
out, that also the
N=2 sector (encoded in the coefficients $\alpha_N^{{\cal T}_\theta}$ and 
$\beta_N^{{\cal T}_\theta}$), which is broken to N=0, shows such a behaviour.
Again, like the cosmological constant (cf. previous section),
\req{gaugeres} takes the same analytic form as the corresponding {\em regularized} 
type $I$ result \cite{ABD}. 

\subsubsection{One--loop Yukawa couplings in theories with broken supersymmetry}
\label{oneloopyukawa}

After having investigated the one--loop cosmological constant and the gauge
threshold corrections in heterotic string vacua with supersymmetry broken by a 
Scherk--Schwarz mechanism, in this section we shall discuss the one--loop Yukawa couplings.

In four--dimensional effective supersymmetric string--theories with canonical 
renormalized Einstein term the 
Yukawa interactions between massless matter fields appear in the form:
\be\label{super}
{\cal L}\sim g_{string}\ e^{\h G^{(0)}}\ W_{ijk}\ \psi^i\psi^j \phi^k+{\rm hc.}\ .
\ee
The function $W_{ijk}$ depends holomorphic on the moduli fields 
and $G^{(0)}$ is the matter and dilaton independent part of the K\"ahler potential.
Thus for the simplest case of one modulus $R$ in a string compactification from five to four
dimensions, we have: $G^{(0)}=-2\ln R$. 
Effective superstring theories are best described by conformal supergravity
with a linear multiplet accounting for the dilaton superfield \cite{jeanpierre}.
The scalar of the compensator is fixed (once) at tree level to get a canonical Einstein term.
The first two additional factors $g_{string} e^{\h G^{(0)}}$ in \req{super} result
from this fixing. Thus they refer to tree--level quantities.
Moreover in this setup, the $F$--density encoding \req{super} 
may not depend on the linear multiplet, which
governs the string--loop expansion. This has the consequence that there is no possibility, 
that \req{super} may receive any perturbative corrections beyond one--loop. 
Any possible renormalization of the Yukawa couplings has to arise from the $D$--density.
More precisely, the physical Yukawa couplings $\lambda_{ijk}$ 
are modified by corrections to the matter field 
K\"ahler metric $Z_{i\ov i}=Z_{i\ov i}^{(0)}+g_{string}^2Z_{i\ov i}^{(1)}+\ldots\ $:
\bea\label{yuk}
\ds{\lambda_{ijk}}&=&\ds{g_{string}\ e^{\h G^{(0)}}\ W_{ijk}\   
(Z_{i\ov i}Z_{j\ov j}Z_{k\ov k})^{-1/2}}\nnn
&\sim& \ds{\  \lambda_{ijk}^{tree}\ 
\lf[1+g_{string}^2\lf(\fc{Z^{(1)}_{i\ov i}}{Z^{(0)}_{i\ov i}}+
\fc{Z^{(1)}_{j\ov j}}{Z^{(0)}_{j\ov j}}+\fc{Z^{(1)}_{k\ov k}}{Z^{(0)}_{k\ov k}}\ri)\ri]^{-1/2}\ .}
\eea
The second line expresses the one--loop approximation.
In particular, to obtain the fermion masses $m_{ij}=\lambda_{ijk}\langle h^k\rangle$, 
one of the metrics $Z_{j\ov j}$ refers to the Higgs field $h$.

The question is how much of the above discussion can be taken over to non--supersymmetric 
effective string theories as they arise e.g. from the model discussed in section \ref{string2}.
The particular feature of toroidal orbifolds with supersymmetry broken by a Scherk--Schwarz 
mechanism is the survival of an N=1 subsector, which is untouched by the 
supersymmetry breaking pattern. Thus its contribution to the effective action should 
still be organized  by N=1 supersymmetry, in particular by the linear multiplet formalism 
reviewed above. Thus Yukawa couplings which only involve matter fields from such an N=1
sector should appear  in the effective action in the form \req{super}. Therefore, by the same arguments as above
the physical Yukawa couplings  $\lambda_{ijk}$ 
between twisted matter fields
should only be corrected by their wave--function renormalization \req{yuk}.

Let us now come to string theory. There, the most canonical way to calculate possible 
one--loop corrections
to Yukawa couplings is to consider the  $S$--matrix 
$\langle V_\psi(z_1)\ V_\psi(z_2)\ V_\phi(z_3)\rangle$
of two fermions and one scalar. Depending on the region of integration $z_i$ this amplitude 
comprises both one--loop vertex corrections $|z_{12}|,|z_{13}|,|z_{23}|>\epsilon$ and one--loop 
propagator corrections (to the external legs) from pinching e.g. $z_1\ra z_2$.
In a supersymmetric string vacuum only the latter contributions give a non--vanishing result 
whereas the one--loop vertex correction vanishes as a simple result of applying
Riemann identities \cite{LL}. This has been explicitly demonstrated for heterotic orbifolds
in ref. \cite{wadia}.
This means, that in agreement to the above field--theoretical arguments Yukawa couplings are not 
renormalized at one--loop in supersymmetric 
string theories\footnote{In fact, the whole holomorphic superpotential $W$ is perturbatively
not renormalized in supersymmetric string vacua \cite{martinec}.}, 
although there are one--loop corrections $Z_{i\ov i}$ to the wave function, modifying 
the physical Yukawa couplings \req{yuk}.
The moduli dependent corrections $Y_i:=Z^{(1)}_{i\ov i}/Z^{(0)}_{i\ov i}$ 
appearing in  eq. \req{yuk} have been calculated for untwisted matter fields
in supersymmetric N=1 orbifold\footnote{The physical Yukawa couplings $\lambda_{ijk}^{tree}$
between three untwisted or one untwisted and two twisted matter fields are constants. 
See refs. \cite{DFMS} for further details on tree--level orbifold couplings.} 
models \cite{AGNT}.
There, the functions $Y_i$ turned out to be proportional to the gauge threshold 
corrections \cite{dkl}.

On the other hand, in generic non--supersymmetric string vacua, the Yukawa string $S$--matrix 
may provide not only non--vanishing propagator corrections $Z^{(1)}_{i\ov i}$ 
but also non--zero vertex corrections $\delta_{ijk}$, 
due to the lack of the relevant Riemann identities.
Therefore, in general the relation \req{yuk} is modified in non--supersymmetric vacua:
\be\label{oneyuk}
\lambda_{ijk}=\lambda^{tree}_{ijk}\ \lf[1-\h g_{string}^2(Y_i+Y_j+Y_k)\ri]
+g_{string}^2\ \delta_{ijk}\ .
\ee
As in the case of gauge threshold corrections (subsection \ref{oneloopgauge})
the quantities $Y_i$ receive radius dependent contributions from the sectors 
${\cal T}_0$ and ${\cal T}_\theta$, only.
Like in the supersymmetric case they take the same radius dependence as the 
one--loop gauge thresholds \req{gaugeres}. It represents a major task, to determine 
$\delta_{ijk}$
for non--supersymmetric string vacua as one has to deal with spin fields arsing in the fermionic 
vertex $V_\psi$ besides string world--sheet instanton corrections at one--loop \cite{atick}.
Though very important, a calculation of $\delta_{ijk}$ is beyond the scope of the present 
article and we leave this for an interesting future project.
However, from the experience made in section \ref{sonntag}, it is  obvious, that
the calculation of $\delta_{ijk}$ must result in an integral of the form \req{Todo0}, which
always gives rise to power suppression in the radius $R$ [cf. eq. \req{Todoi}].


\sect{Conclusions}
The paper investigated the properties of the scalar potential/vacuum
energy in effective field theory models with one additional
compact dimension,   as well as in the 
heterotic and type I string compactifications.

In the effective field theory approach there are
additional  KK states associated
with the extra dimension which affect significantly the scalar potential.
The limit of summing the whole Kaluza Klein tower was
investigated to show the absence at one loop of quadratic and logarithmic
divergences in either the bosonic or fermionic  sector, respectively.
In this mechanism the presence of states of mass larger than
the cut-off the 4D effective theory  played an important role \cite{dmg}.
This prompted us to investigate this problem in the context of string theory.

While the scalar potential/Higgs mass receives a finite 
(Yukawa) contribution in the ``KK regularization''
limit at one loop level, the dependence of the perturbative expansion on the
Yukawa or gauge couplings may make this result ultraviolet sensitive
at two loop level and beyond. This dependence is due to
the fact that the ``running''  gauge and Yukawa couplings  
are ultraviolet sensitive in one loop and higher orders.
Gauge corrections can bring a quadratic divergence even at
one loop level \cite{toappear}.
The non-renormalizable character of Kaluza Klein models
may further enhance this ultraviolet sensitivity.

In section 3 we derived formulas for the one--loop cosmological constant \req{Generic} 
and gauge couplings \req{gaugeres}
for generic $D=4$ string compactifications with supersymmetry broken
by a Scherk--Schwarz mechanism w.r.t to one internal circle of radius $R$.
In particular, the one--loop gauge couplings show a logarithmic
dependence \req{gaugeres}
on the compactification scale $R$ in contrast to a linear $R$ 
behaviour in the corresponding  supersymmetric case \req{nyc}.
Our results, also relevant for $D=5$ supersymmetric string theories
at finite temperature $T=1/(2\pi R)$, can be nicely rewritten in 
terms of polylogarithms
appropriate for a comparison with field--theory (cf. \req{polylog},
\req{Todoi} and (\ref{vfinale}), (\ref{aa2})). 
We compared our heterotic string calculations, which are $UV$ finite due to 
the power of world--sheet modular invariance or/and the interplay 
between KK and winding
states, with corresponding type $I$ calculations. In general, the
latter need to be regularized like ordinary field--theories. However, 
their finite part assumes the same analytic form \req{clau} as the 
heterotic results. 
This correspondence at one--loop may be a remnant of heterotic--type 
$I$ duality in nine dimensions. 

Arguments in favor of a full string calculation were presented
in the framework of general Kaluza Klein (orbifold) models.
It has been shown that the limit of summing the 
{\it whole} Kaluza Klein tower provides a result compatible with
a {\it full} (heterotic) string result which actually includes not only 
the effects of KK
states, but also the effects due to  winding modes as well, whose 
presence cannot be explained and is not accounted for 
by current  field theory calculations or ``KK regularization''.
An explanation why these two results are equal 
has been suggested in Section 2.5  as due to 
a discrete ``shift symmetry'' in the absence of winding modes
(see \cite{stephan2}), as a remnant of modular
invariance in (heterotic) string theory.
Further, the comparison of the effects of KK states alone in the effective 
field theory and
string theory shows that they have different structure, and that  the 
field theory limit of the string result  
requires a truncation of the KK tower to match it onto an effective
field theory result with  Kaluza Klein states of mass  
below some cut-off of the order of $M_{s}$.
In this way the heterotic string can provide a physical regularization
scheme of the 4D effective field theory result with a truncated tower
of Kaluza Klein states.

From this analysis we conclude that radiative corrections generically
tend to raise the value of the relevant physical quantities (as e.g. 
Higgs mass) towards the string scale $M_{string}$.
Unless protected by a symmetry like low energy softly broken
supersymmetry, the ultraviolet sensitivity of physical quantities
will be manifest, even in the case of a large extra dimension 
$1/R\ll M_{string}$. 

\vspace{2cm} 
\noindent
{\bf Acknowledgments}\\

\noindent
We would like to thank C. Burgess, S. Groot Nibbelink and T. Taylor 
for interesting and helpful discussions.\\
The work of D.M.G. and H.P.N. was supported by the 
University of Bonn under the European Commission RTN programme
HPRN-CT-2000-00131 and 00148.
St. St. is supported by the National Science Foundation 
under grant PHY--99--01057.

\newpage

\section*{Appendix}
\appendix
\sect{Kaluza--Klein sums}
We have the following mathematical  identities 
\begin{equation}\label{sumtoinf}
\sum_{k=-\infty}^{\infty}
\ln\frac{\rho^2+\pi^2( k+\omega')^2}{\rho^2+\pi^2( k+\omega)^2}=
\ln\frac{\cosh(2 \rho)-\cos(2\pi\omega')}
{\cosh(2 \rho)-\cos(2\pi\omega)}
\end{equation}
This relation can easily 
be proved using the following mathematical identity \cite{gradshteyn}:
\begin{equation}\label{cosh}
\cosh(2 \rho)-\cos(2\pi\omega')=2 \sin^2(\pi\omega')
\prod_{k=-\infty}^{\infty}
\left(1+\frac{\rho^2}{\pi^2(k+\omega')^2}\right)
\end{equation}
We take the logarithm of this last equation which we then write  for both 
$\omega$ and $\omega'$  and  subtract the two results to 
reconstruct the left hand side of equation (\ref{sumtoinf}).
Further we make use of the following identity \cite{gradshteyn}:
\begin{equation}
\frac{\sin \pi\omega'}{\sin \pi \omega}=\prod_{k=-\infty}^{\infty}\left(
1+\frac{\omega'-\omega}{k+\omega}\right)
\end{equation}
We take the logarithm of this expression and after some algebra
we can easily recover from the last two equations,
 the result of eq.(\ref{sumtoinf}).

In the text we also made use of the following sum of the effects of a 
truncated tower of Kaluza-Klein states, which is proved below
\begin{equation}\label{sumtot}
\sum_{k=-l}^{l}\ln\frac{\rho^2+\pi^2( k+\omega')^2}
{\rho^2+\pi^2( k+\omega)^2}={\cal Z}_0+{\cal Z}_1+{\cal Z}_2
\end{equation}
with the notation
\begin{equation}\label{zz0}
{\cal Z}_0=\ln\frac{\cosh(2 \rho)-\cos(2\pi\omega')}
{\cosh(2 \rho)-\cos(2\pi\omega)}
\end{equation}
and
\begin{equation}\label{zz1}
{\cal Z}_1=
\ln\frac{[\rho^2+\pi^2( l\pm\omega')^2]_*}
{[\rho^2+\pi^2( l\pm\omega)^2]_*}
\end{equation}
and finally
\begin{equation}\label{zz2}
{\cal Z}_2=
\ln\frac{[\Gamma(l \pm \omega'\pm i\rho/\pi)]_*}
{[\Gamma(l \pm \omega\pm i\rho/\pi)]_*}
\end{equation}
In these equations 
the symbol $[\rho^2+\pi^2( l\pm\omega)^2]_*$ means that a product
of the expression within $[..]_*$  with all combinations of plus and minus
signs is considered (similar for $[\Gamma[(l \pm \omega'\pm
i\rho/\pi)]_*$)). We observe that taking the limit 
$l\rightarrow \infty$ should  recover the result of eq.(\ref{sumtoinf}).
Since the second term (${\cal Z}_1$) in (\ref{sumtol}) has vanishing limit
($l\rightarrow \infty$, for fixed
$\rho$) we conclude that the last term of (\ref{sumtol}) (${\cal Z}_2$)
has in such a case a  vanishing limit as well.
The integral of ${\cal Z}_0$,  ${\cal Z}_1$ and ${\cal Z}_2$
with respect to $\rho$
leads to the potentials $\Vc_0$, $\Vc_1$ and $\Vc_2$ respectively,
eqs.(\ref{pot0}),(\ref{pot1}),(\ref{pot2}) in the text.
For the aforementioned reasons, in the limit of large $l$ and fixed
$\overline\Lambda$, the potentials $\Vc_1$ and $\Vc_2$ vanish.

To prove eq.(\ref{sumtot})  we start with the following 
mathematical identity
\begin{eqnarray}\label{in}
\sum_{k=-l}^{l}\ln(a+\pi(k+\omega'))&=&-\ln(a+\pi\omega' )
+\ln\Gamma(1+l-\omega'-a/\pi)+\ln\Gamma(1+l+\omega'+a/\pi)\nonumber\\
&&-\ln\Gamma(\omega'+a/\pi)-\ln\Gamma(-\omega'-a/\pi)
\end{eqnarray}
We write this equation  with the replacement $a\rightarrow \pm i\rho$;
then we  re-write the obtained result  
for the case with  $\omega'\rightarrow \omega$; we  then  subtract
the two results to reconstruct the truncated sum of the left
hand side of (\ref{sumtot}). 
We find that
\begin{eqnarray}\label{truncated}
\sum_{k=-l}^{l}\ln\frac{\rho^2+\pi^2( k+\omega')^2}
{\rho^2+\pi^2( k+\omega)^2}&=&
-\ln\frac{\rho^2+\pi^2 \omega'^2}{\rho^2+\pi^2\omega^2}
+\ln\frac{[\rho^2+\pi^2( l\pm\omega')^2]_*}
{[\rho^2+\pi^2( l\pm\omega)^2]_*}
\nonumber\\
&&-\ln\frac{[\Gamma(\pm \omega'\pm i\rho/\pi)]_*}
{[\Gamma(\pm \omega\pm i\rho/\pi)]_*}
+\ln\frac{[\Gamma(l \pm \omega'\pm i\rho/\pi)]_*}
{[\Gamma(l \pm \omega\pm i\rho/\pi)]_*}
\end{eqnarray}
where for the terms $\Gamma(1+l\pm\omega'\pm i\rho/\pi)$
we made use of the relation
\begin{equation}
\Gamma[1+x]=x \Gamma[x]
\end{equation}
In eq.(\ref{truncated})
we apply the identity written below \cite{gradshteyn}
\begin{equation}\label{product}
\frac{\Gamma(\alpha) \Gamma(\beta)}
{\Gamma(\alpha+\gamma)\Gamma(\beta-\gamma)}
=\prod_{k=0}^{\infty}\left(1+\frac{\gamma}{k+\alpha}\right)
\left(1-\frac{\gamma}{k+\beta}\right)
\end{equation}
More explicitly we  consider the logarithm  of this 
equality for the case  ($\alpha,\beta\rightarrow \omega'$, 
$\gamma\rightarrow i\rho/\pi$)
and also for the case ($\alpha,\beta\rightarrow - \omega'$,
$\gamma\rightarrow i\rho/\pi$). This will provide us an expression for
the third term in (\ref{truncated}),  $\ln[\Gamma(\pm \omega'\pm
i\rho/\pi)]_*$. We further  replace the products
of $\Gamma$ functions of arguments of opposite signs
$\Gamma(-\omega')\Gamma(\omega')$
by the identity \cite{gradshteyn}
\begin{equation}
\Gamma[\omega']\Gamma[-\omega']=-\frac{\pi}{\omega' \sin\pi \omega'}
\end{equation} 
Finally in the result for the truncated sum (\ref{truncated})
we replace the infinite series of (\ref{product}) with the left hand
side of equation (\ref{cosh}) to obtain the desired result, 
eq.(\ref{sumtot}).\par

\section{Asymptotic expansions for the  potential}

In this section we study the asymptotic behaviour
of the product of Gamma functions 
\begin{equation}\label{as}
\ln\frac{[\Gamma(l \pm \omega'\pm i\rho/\pi)]_*}
{[\Gamma(l \pm \omega\pm i\rho/\pi)]_*}
\end{equation}
when their  arguments have a large modulus, 
i.e. when $|l+i \rho/\pi|$ is very large, without specifying 
the relationship between $l$ and $\rho$.
For this we use the following expansion valid for large $|z|$ 
\cite{gradshteyn} 
\begin{equation}
\ln\Gamma(z)\approx z \ln z-z -\frac{1}{2} \ln z+\frac{1}{2}\ln(2\pi)
+\frac{1}{12 z}+{\cal O}(z^{-3})
\end{equation}
We therefore apply this equation by replacing $z$ with the various
arguments of Gamma functions present in (\ref{as}).
After some algebra we find that
\begin{eqnarray}\label{piece}
\ln\frac{[\Gamma(l \pm \omega'\pm i\rho/\pi)]_*}
{[\Gamma(l \pm \omega\pm i\rho/\pi)]_*}&\approx &
\left\{l-\frac{1}{2}\right\}
\ln\frac{[\rho^2+\pi^2( l\pm\omega')^2]_*}
{[\rho^2+\pi^2( l\pm\omega)^2]_*}\nonumber\\
&&+
\left\{\frac{i\rho}{\pi}\ln 
\frac{[\pi(l\pm\omega')+i\rho]_*}{[\pi(l\pm\omega')-i\rho]_*}
+\omega'\ln
\frac{\pi^2(l+\omega')^2+\rho^2}{\pi^2(l-\omega')^2+\rho^2}
-(\omega'\rightarrow \omega)\right\}\nonumber\\
&&+\frac{\pi^2}{6}\left\{\frac{l-\omega'}
{\rho^2+\pi^2(l-\omega')^2}
+\frac{l+\omega'}
{\rho^2+\pi^2(l+\omega')^2}-(\omega'\rightarrow \omega)\right\}
\end{eqnarray}
This  expression is then integrated piecewise to provide an 
asymptotic behaviour of $\Vc_2(\phi)$ for large
value of the absolute value of the complex number $l+i\overline
{\Lambda}$, without a restriction on the relative behaviour
of $l$ and ${\overline\Lambda}$. 
The first term in (\ref{piece}), 
when integrated gives a contribution proportional
to $\Vc_1$, so a cancellation of  
quadratic and logarithmic divergences 
as noticed for $\Vc_1$ applies in this case, too.
For each of the other two  terms a similar cancellation 
takes place.

\sect{Kaluza--Klein Integrals}

In this section we provide useful mathematical 
formulas needed to evaluate various  Kaluza Klein
integrals. Their asymptotic behaviour will also be carefully
investigated.

We would like to investigate the behaviour of the following 
Kaluza Klein integrals, for $\alpha=-2$ which where encountered
in the text, while evaluating the vacuum energy.
\begin{equation}
{\cal I}(\epsilon^2,\alpha)= \frac{1}{2}\int_{\epsilon^2}^{\infty}
dt\, t^{\alpha-1} \sum_{k=-\infty}^{\infty}e^{-k^2 \pi t}
\end{equation}
and 
\begin{equation}
{\cal J}(\epsilon^2,\alpha)= \frac{1}{2}\int_{\epsilon^2}^{\infty}
dt\, t^{\alpha-1} \sum_{k=-\infty}^{\infty}e^{-(k+1/2)^2 \pi t}
\end{equation}
As a first step we provide a useful integral representation of 
Riemann zeta function \cite{gradshteyn} valid for {\it all} 
complex/real values of its argument, $\alpha$.
\begin{equation}
\zeta(2\alpha)=\frac{\pi^\alpha}{\Gamma(\alpha)}
\left\{\frac{1}{2\alpha (2\alpha-1)}+{\cal K}\right\}, \,\,\,\,\,\,\,\,
{\cal K}=\int_{1}^{\infty}dt \,\left(t^{-1/2-\alpha}+t^{\alpha-1}\right)
\sum_{k=1}^{\infty}e^{-k^2 \pi t}
\end{equation}
It is useful for our purposes to transform  the integral 
of ${\cal K}$ into a Kaluza-Klein integral, with lower limit of
integration equal to $\epsilon^2\ra 0$.
\begin{eqnarray}
{\cal K(\alpha)}&=&\int_{1}^{\infty}dt\, t^{-1/2-\alpha}
\sum_{k=1}^{\infty}e^{-k^2 \pi t}+
\int_{1}^{\infty}dt\, t^{\alpha-1}
\sum_{k=1}^{\infty}e^{-k^2 \pi t}\nn\\
&=&
\int_{\epsilon^2}^{1}dt\, t^{-3/2+\alpha}
\sum_{k=1}^{\infty}e^{-k^2 \pi /t}+
\int_{1}^{\infty}dt\, t^{\alpha-1}
\sum_{k=1}^{\infty}e^{-k^2 \pi t}\nn\\
&=&\frac{1}{2}\int_{\epsilon^2}^{1}dt \, t^{-3/2+\alpha}
\left[\sum_{k=-\infty}^{\infty} e^{-k^2 \pi /t}-1\right]
+\int_{1}^{\infty}dt\, t^{\alpha-1}
\sum_{k=1}^{\infty}e^{-k^2 \pi t}\nn\\
&=&\frac{1}{2}\int_{\epsilon^2}^{1}dt \, t^{-3/2+\alpha}
\left[t^{1/2} \sum_{k=-\infty}^{\infty}e^{-k^2 \pi t}-1\right]
+\frac{1}{2}\int_{1}^{\infty}dt\, t^{\alpha-1}
\left[\sum_{k=-\infty}^{\infty}e^{-k^2 \pi t}-{1}\right]\nn\\
&=&{\cal I}(\epsilon^2,\alpha)-
\frac{1}{2}\int_{\epsilon^2}^{1}dt\, t^{-3/2+\alpha}
-\frac{1}{2}\int_{1}^{\infty}dt\,t^{\alpha-1}\label{K}
\end{eqnarray}
where in the last step a Poisson resummation over $k$
has been performed under the first integral.
We therefore find the following behaviour of 
${\cal I}(\epsilon^2,\alpha=-2)$
\begin{equation}
{\cal I}(\epsilon^2\ra 0,\alpha=-2)\equiv\frac{1}{2}\int_{\epsilon^2}^{\infty}
\frac{dt}{t^{3}} \sum_{k=-\infty}^{\infty}e^{-k^2 \pi t}
=\frac{3}{4\pi^2}\zeta(5)+\frac{1}{5}
\epsilon^{-5}
\end{equation}
To evaluate ${\cal J}(\epsilon^2,\alpha)$ we split its Kaluza Klein sum
over odd and even $k$ to find that
\begin{eqnarray}
{\cal I}(\epsilon^2,\alpha)&=& \frac{1}{2}\int_{\epsilon^2}^{\infty}
dt\, t^{\alpha-1} \sum_{k=-\infty}^{\infty}e^{-k^2 \pi t}\nn\\
&=&
\frac{1}{2}\int_{\epsilon^2}^{\infty}
dt\, t^{\alpha-1} \sum_{n=-\infty}^{\infty}e^{-4 n^2 \pi t}+
\frac{1}{2}\int_{\epsilon^2}^{\infty}
dt\, t^{\alpha-1} \sum_{n=-\infty}^{\infty}e^{-4(n+1/2)^2 \pi t}
\end{eqnarray}
which after a rescaling of the integration variable $t$, 
$ t\rightarrow t'=4 t$ gives
\begin{equation}
{\cal I}(\epsilon^2,\alpha)=4^{-\alpha} {\cal I}(4 \epsilon^2,\alpha)
+4^{-\alpha} {\cal J}(4 \epsilon^2,\alpha)
\end{equation}
Using the expression of ${\cal I}(\epsilon^2,\alpha=-2)$
we find that
\begin{equation}
{\cal J}(\epsilon^2\ra 0,\alpha=-2)\equiv\frac{1}{2}\int_{\epsilon^2}^{\infty}
\frac{dt}{t^{3}} \sum_{k=-\infty}^{\infty}e^{-(k+1/2)^2 \pi t}=
\frac{-15}{16} \frac{3}{4\pi^2}\zeta(5)+\frac{1}{5}
\epsilon^{-5}
\end{equation}

We therefore find the following results for fermionic and bosonic  
Kaluza Klein integrals used in the literature:
\begin{equation}
\frac{1}{2}\int_{\epsilon^2 R^2}^{\infty}
\frac{dt}{t^{3}} \sum_{k=-\infty}^{\infty}e^{-k^2 \pi t/R^2}
=R^{-4}{\cal I}(\epsilon^2,\alpha=-2)=
\frac{1}{R^{4}}\left\{
\frac{3}{4\pi^2}\zeta(5)+\frac{1}{5}
\epsilon^{-5}\right\}
\end{equation}
and also
\begin{equation}
\frac{1}{2}\int_{\epsilon^2 R^2}^{\infty}
\frac{dt}{t^{3}} \sum_{k=-\infty}^{\infty}e^{-(k+1/2)^2 \pi t/R^2}
=R^{-4}{\cal J}(\epsilon^2,\alpha=-2)=
\frac{1}{R^4}\left\{\frac{-15}{16} 
\frac{3}{4\pi^2}\zeta(5)+\frac{1}{5}
\epsilon^{-5}\right\}
\end{equation}
which shows that the two integrals above have same type of divergence, 
in $\epsilon^2=0$ with equal coefficient multiplying it and canceling
in their difference. The origin
of this divergence for $(\alpha=-2)$ can be traced back 
to the $k=0$ Kaluza Klein mode 
added and subtracted under the first integral in the third line of
eq.(\ref{K}). 
In terms of a dimensionful lower limit of integration,
$\varepsilon^2=\epsilon^2 R^2$, $\varepsilon\propto (mass)^{-1}$
we find
\begin{equation} \label{oneint}
\frac{1}{2}\int_{\varepsilon^2}^{\infty}
\frac{dt}{t^{3}} \sum_{k=-\infty}^{\infty}e^{-k^2 \pi t/R^2}
={R^{-4}}{\cal I}(\varepsilon^2/R^2,\alpha=-2)=
\frac{1}{R^{4}}
\frac{3}{4\pi^2}\zeta(5)+\frac{1}{5}
\varepsilon^{-5} R
\end{equation}
and also
\begin{equation}\label{twoint}
\frac{1}{2}\int_{\varepsilon^2}^{\infty}
\frac{dt}{t^{3}} \sum_{k=-\infty}^{\infty}e^{-(k+1/2)^2 \pi t/R^2}
=R^{-4}{\cal J}(\varepsilon^2/R^2,\alpha=-2)=
\frac{1}{R^4}\frac{-15}{16} 
\frac{3}{4\pi^2}\zeta(5)+\frac{1}{5}
\varepsilon^{-5} R
\end{equation}
or
\begin{equation}\label{twin}
\frac{1}{2}\int_{\varepsilon^2}^{\infty}
\frac{dt}{t^{3}} \sum_{k=-\infty}^{\infty}
\left[e^{-k^2 \pi t/R^2}-
e^{-(k+1/2)^2 \pi t/R^2}\right]
=
\frac{1}{R^4}\frac{31}{16} 
\frac{3}{4\pi^2}\zeta(5)
\end{equation}
encountered in the text.
The conclusion is that neither ${\cal I}(\varepsilon^2/R^2,\alpha=-2)$
nor ${\cal J}(\varepsilon^2/R^2,\alpha=-2)$ has any quadratic or logarithmic
divergences after summing up an infinite  tower of states.

\vspace{1.3cm}
Using the same procedure we also find the following result for the
integrals used in Section~\ref{sonntag}, eq.(\ref{Todo}) for n=2, D=4, N=0.
\begin{equation}
{\cal I}(\epsilon^2,\alpha=0,\Lambda)\equiv \frac{1}{2}
\int_{\epsilon^2}^{\Lambda}
\frac{dt}{t} \sum_{k=-\infty}^{\infty}e^{-k^2 \pi t}
=\frac{1}{2}\gamma_E-\frac{1}{2}
\ln(4\pi)+\epsilon^{-1}+\frac{1}{2}\ln \Lambda, \,\,\,\Lambda\ra\infty
\end{equation}
and 
\begin{equation}
{\cal J}(\epsilon^2,\alpha=0,\Lambda)\equiv \frac{1}{2}
\int_{\epsilon^2}^{\Lambda}
\frac{dt}{t} \sum_{k=-\infty}^{\infty}e^{-(k+1/2)^2 \pi t}
=\epsilon^{-1}-\ln 2
\end{equation}
Therefore the divergence $\epsilon^{-1}$ will cancel in the difference
of the two integrals which  is equal to (for R finite) 
\begin{equation}\label{eqA}
{\cal P}\equiv \frac{1}{2}
\int_{\epsilon^2 R^2\ra 0}^{\Lambda\ra\infty} \frac{dt}{t}  
\sum_{k=-\infty}^{\infty}
\left[e^{-k^2 \pi t/R^2}-e^{-(k+1/2)^2 \pi t/R^2}\right]
=\frac{1}{2}\left(\gamma_E-\ln\pi+\ln \Lambda-\ln R^2\right)
\end{equation}
Note that ${\cal P}$ is the Poisson re-summed form of equation
(\ref{Todo}). Indeed
\be
{\cal P}=R \int_{\epsilon^2 R^2}^{\infty} \frac{dt}{t^{3/2}}
\sum_{n\geq 0}^{}
\left[1-(-1)^n\right] e^{- n^2 \pi R^2/t}
\ee
Note that the divergent part $\ln \Lambda$  is removed from the 
expression of ${\cal P}$, eq.(\ref{eqA}) by the massless modes 
accounted for by the ``-1'' term subtracted from the partition 
function under the integral over the fundamental domain which 
gives the general formula for the gauge thresholds \cite{dkl}. 
For this reason $\ln \Lambda$ is not present in the
final result, last line of eq.(\ref{Todoi}).

\sect{Lattice functions}

It is convenient to introduce the lattice functions \cite{Rohm}:
\bea
\ds{\Ec_0}&=&\ds{\sum_{m \in Z\atop n \in 2\IZ}\ e^{2\pi i \tau_1 mn}\ 
e^{-\pi\tau_2[\fc{m^2}{R^2}+n^2R^2]}\ ,}\nnn
\ds{\Ec_{1/2}}&=&\ds{\sum_{m \in Z\atop n \in 2\IZ}\ e^{2\pi i \tau_1 (m+\h)n}\ 
e^{-\pi\tau_2[\fc{(m+\h)^2}{R^2}+n^2R^2]}\ ,}\nnn
\ds{\Oc_0}&=&\ds{\sum_{m \in Z\atop n \in 2\IZ}\ e^{2\pi i \tau_1 m(n+1)}\ 
e^{-\pi\tau_2[\fc{m^2}{R^2}+(n+1)^2R^2]}\ ,}\nnn
\ds{\Oc_{1/2}}&=&\ds{\sum_{m \in Z\atop n \in 2\IZ}\ e^{2\pi i \tau_1 (m+\h)(n+1)}\ 
e^{-\pi\tau_2[\fc{(m+\h)^2}{R^2}+(n+1)^2R^2]}\ .}\nnn
\label{rohmfunctions}
\eea
Poisson resummation on \req{rohmfunctions} leads:
\bea
\ds{\Ec_0}&=&\ds{\fc{R}{\tau_2^{1/2}}\sum_{n,l\in \IZ}\ 
[1+(-1)^n]\ e^{-\fc{\pi R^2}{\tau_2}|l+n\tau|^2}\ ,}\nnn
\ds{\Ec_{1/2}}&=&\ds{\fc{R}{\tau_2^{1/2}}\sum_{n,l\in \IZ}\ 
[1+(-1)^n](-1)^l\ e^{-\fc{\pi R^2}{\tau_2}|l+n\tau|^2}\ ,}\nnn
\ds{\Oc_0}&=&\ds{\fc{R}{\tau_2^{1/2}}\sum_{n,l\in \IZ}\ 
[1-(-1)^n]\ e^{-\fc{\pi R^2}{\tau_2}|l+n\tau|^2}\ ,}\nnn
\ds{\Oc_{1/2}}&=&\ds{\fc{R}{\tau_2^{1/2}}\sum_{n,l\in \IZ}\ 
[1-(-1)^n]\ (-1)^l\ e^{-\fc{\pi R^2}{\tau_2}|l+n\tau|^2}\ .}\nnn
\label{rohmfunctionsii}
\eea
The modular properties for these functions can be easily deduced:
\bea
\ds{\Ec_0(-1/\tau)}    &=&  \ds{\h|\tau|\ (\Ec_0+\Ec_{1/2}+\Oc_0+\Oc_{1/2})\ ,}\nnn
\ds{\Ec_{1/2}(-1/\tau)}&=&  \ds{\h|\tau|\ (\Ec_0+\Ec_{1/2}-\Oc_0-\Oc_{1/2})\ ,}\nnn
\ds{\Oc_0(-1/\tau)}    &=&  \ds{\h|\tau|\ (\Ec_0-\Ec_{1/2}+\Oc_0-\Oc_{1/2})\ ,}\nnn
\ds{\Oc_{1/2}(-1/\tau)}&=&  \ds{\h|\tau|\ (\Ec_0-\Ec_{1/2}-\Oc_0+\Oc_{1/2})\ ,}\nnn
\label{rohmfunctionsiii}
\eea
and $\Ec_0(\tau+1)=\Ec_0,\ \Ec_{1/2}(\tau+1)=\Ec_{1/2},\ \Oc_0(\tau+1)=\Oc_0$ and 
$\Oc_{1/2}(\tau+1)=-\Oc_{1/2}$.

\sect{Modular functions for gauge thresholds}
\def\qb{\ov q}
\def\th{\theta}
In this appendix we present the modular functions needed for 
the gauge threshold calculation
in section \ref{sonntag} of the model in section \ref{string2}.
For the sector ${\cal T}_0$ we have\footnote{The arrow symbolizes projection onto equal 
left/right masses, which appears after the $\tau_1$--integration. Cf. also \req{numbers}.}  :
\bea
\ds{{\cal A}(q,\qb)}&=&\ds{\fc{1}{72}\fc{E_4(q)}{\eta(\ov q)^{12}\eta(q)^{24}}
[\theta_3(\ov q)^4+\th_4(\qb)^4+E_2(\qb)]\ [E_4(q)E_2(q)-E_6(q)]\ 
[{\cal N}_{R=1}(q,\qb)]^5}\nnn
&\ra&\ds{\sum_{N>0}\ \alpha^{{\cal T}_0}_N\ q^N\ \qb^N\ ,}\nnn
&=&\ds{480+9600\ q^{1/4}\qb^{1/4}+76800\ q^{1/2}\qb^{1/2}+
307200\ q^{3/4}\qb^{3/4}+\ldots\ ,}\nnn
\ds{{\cal B}(q,\qb)}&=&\ds{\fc{1}{72}\fc{E_4(q)}{\eta(\ov q)^{12}\eta(q)^{24}}
\lf\{ [\theta_3(\ov q)^4+\th_4(\qb)^4+E_2(\qb)] E_4(q)+E_4(q)E_2(q)-E_6(q)\ri\}
[{\cal N}_{R=1}(q,\qb)]^5}\nnn
&\ra&\ds{\sum_{N>0}\ \beta^{{\cal T}_0}_N\ q^N\ \qb^N}\nnn
&=&\ds{\fc{1}{3}\ (\ 1508+30080\  q^{1/4}\qb^{1/4}+240000\ q^{1/2}\qb^{1/2}+957440\ 
q^{3/4}\qb^{3/4}+\ldots\ )\ ,}\nnn

\eea
Of course, before the supersymmetry breaking, we have
a vanishing $\beta$--function coefficient in the N=4 sector 
\be
b^{{\cal T}_0}_{E_8'}=0
\ee
and after introducing the Scherk--Schwarz shifts \req{shifts} we obtain:
\be
b^{{\cal T}_0}_{E_8'}=480\equiv \alpha^{{\cal T}_0}_0\ .
\ee
For the sector ${\cal T}_\theta$ we have the modular functions:
\bea
\ds{{\cal A}(q,\qb)}&=&\ds{-\fc{1}{72}\ \fc{1}{\eta(\qb)^6\eta(q)^{18}}\ 
[\theta_3(\ov q)^4+\th_4(\qb)^4+E_2(\qb)]\ {\cal N}_{R=1}(q,\qb)}\nnn
&\times&\ds{[E_4(q)E_2(q)-E_6(q)]\ 
\lf\{\lf|\fc{\th_2^2\th_3^2}{\th_4^2}\ri|^2[\th_3(q)^4+\th_2(q)^4]+
\lf|\fc{\th_2^2\th_4^2}{\th_3^2}\ri|^2[\th_2(q)^4-\th_4(q)^4]\ri\}}\nnn
&\ra&\ds{\sum_{N>0}\ 
\alpha^{{\cal T}_\theta}_N\ q^N\ \qb^N\ ,}\nnn
&=&\ds{-7680\ q^{1/2}\ \qb^{1/2}-30720\ q^{3/4}\ \qb^{3/4}-1536000\ q\ \qb
-5898240\ q^{5/4}\qb^{5/4}\ldots}
\eea
Now in the N=2 sector alone, before supersymmetry breaking we encounter the usual
$E_8'$ $\beta$--function coefficient for standard gauge 
embedding (see e.g.: \cite{gauge2}) 
\be
b^{{\cal T}_\theta}_{E_8'}=60
\ee
and after introducing the Scherk--Schwarz shifts \req{shifts} we determine:
\be
b^{{\cal T}_\theta}_{E_8'}=-30\ .
\ee
\bea
\ds{{\cal B}(q,\qb)}&=&\ds{-\fc{1}{72}\ \fc{1}{\eta(\qb)^6\eta(q)^{18}}\ 
\lf\{\lf|\fc{\th_2^2\th_3^2}{\th_4^2}\ri|^2[\th_3(q)^4+\th_2(q)^4]+
\lf|\fc{\th_2^2\th_4^2}{\th_3^2}\ri|^2[\th_2(q)^4-\th_4(q)^4]\ri\} }\nnn
&\times&\ds{\lf\{E_4(q)E_2(q)-E_6(q)
+[\theta_3(\ov q)^4+\th_4(\qb)^4+E_2(\qb)]E_4(q)\ri\}
\ {\cal N}_{R=1}(q,\qb)}\nnn
&\ra&\ds{\sum_{N>0}\ 
\beta^{{\cal T}_\theta}_N\ q^N\ \qb^N\ ,}\nnn
&=&\ds{-\fc{1}{3}\ (\ 128+512\ q^{1/4}\qb^{1/4}+24000\ q^{1/2}\ \qb^{1/2}+
95744\ q^{3/4}\ \qb^{3/4}+3197440\ q\ \qb+\ \ldots\ )}
\eea

\sect{Partition functions for Wilson line dependence}

In this section we want to present the partition functions needed in section 
\ref{wilsondependence} 
for the model of \ref{string2} when the Wilson line \req{wilson} is turned on.

Following the explanations in section \ref{wilsondependence} the 
partition function \req{N=4cosmo} of the N=4 sector ${\cal T}_0$
is modified to: 
\be
\sum_{M>-1\atop N\geq 0}\ \gamma^{{\cal T}_0}(M,N)\ q^M\ \ov q^N=
\eta^{-12}(\ov q)\eta^{-24}(q)\ \theta_2(\ov q)^4
E_4(q)\ [E_{4,1}^{even}(q)+E_{4,1}^{odd}(q)]\ \Nc_{SO(5,5)}(q,\ov q)\ .
\ee
Furthermore, for the N=2 sector ${\cal T}_\theta$ we obtain 
\be
\sum_{M>-1/2\atop N\geq 0}\ \gamma^{{\cal T}_\theta}(M,N)\ q^M\ \ov q^N=
-\fc{E_{4,1}^{even}(q)+E_{4,1}^{odd}(q)}{\eta^{18}\ov\eta^6}
\lf[\lf|\fc{\theta_2^2\theta_3^2}{\theta_4^2}\ri|^2 (\theta_3^4+\theta_2^4)+
\lf|\fc{\theta_2^2\theta_4^2}{\theta_3^2}\ri|^2 (-\theta_4^4+\theta_2^4)\ri]\ .
\ee
The combination of both coefficients $\gamma^{{\cal T}_0}(M,N),\ \gamma^{{\cal T}_\theta}(M,N)$
gives the total contribution\\ $\gamma(M,N)=\fc{1}{4}\gamma^{{\cal T}_0}(M,N)+
\fc{1}{4}\gamma^{{\cal T}_\theta}(M,N)$, needed in  \req{wlresult}.



\end{document}